\documentclass[prd,floats,preprint,superscriptaddress,floatfix,eqsecnum,nofootinbib,12pt]{revtex4}
\pdfoutput=1

\usepackage{amsmath,amssymb,graphicx,bigstrut,ulem,epsfig,tensor,hyperref,color,xcolor}

\makeatletter
\DeclareRobustCommand{\em}{%
  \@nomath\em \if b\expandafter\@car\f@series\@nil
  \normalfont \else \itshape \fi}
\makeatother

\usepackage[footnotesize,bf,margin=.5in,flushleft,compatibility=false]{caption}
\usepackage{subcaption}
\usepackage{braket}
\usepackage{tikz}
\usetikzlibrary{decorations.markings,decorations.pathmorphing}
\tikzset{->-/.style={decoration={markings,mark=at position #1 with {\arrow{>}}},postaction={decorate}}}
\tikzset{>=latex}

\newcommand{\D}{\ensuremath{\mathrm{d}}}
\newcommand{\cD}{\ensuremath{\mathcal{D}}}

\newcommand{\order}[1]{\ensuremath{\mathcal{O}\left( #1 \right)}}
\renewcommand{\Re}{\ensuremath{\text{Re}}}
\renewcommand{\Im}{\ensuremath{\text{Im}}}
\newcommand{\cP}{\ensuremath{\mathcal{P}}}
\newcommand{\fx}{\ensuremath{\mathbf{x}}}
\newcommand{\fX}{\ensuremath{\mathbf{X}}}
\newcommand{\tr}{\ensuremath{\text{Tr}}}

\def\be{\begin{equation}}
\def\ee{\end{equation}}

\begin{document}

\title{Lorentzian Condition in Holographic Cosmology}

\author{Thomas Hertog}
\email{thomas.hertog@kuleuven.be}
\affiliation{Institute for Theoretical Physics, KU Leuven, 3001 Leuven, Belgium}
\author{Ruben Monten}
\email{ruben.monten@kuleuven.be}
\affiliation{Institute for Theoretical Physics, KU Leuven, 3001 Leuven, Belgium}
\affiliation{Department of Physics, Columbia University, 538 West 120th Street, New York, New York 10027}
\author{Yannick Vreys}
\email{yannick.vreys@kuleuven.be}
\affiliation{Institute for Theoretical Physics, KU Leuven, 3001 Leuven, Belgium}
\begin{abstract}

We derive a sufficient set of conditions on the Euclidean boundary theory in dS/CFT for it to predict classical, Lorentzian bulk evolution at large spatial volumes. Our derivation makes use of a canonical transformation to express the bulk wave function at large volume in terms of the sources of the dual partition function. This enables a sharper formulation of dS/CFT. The conditions under which the boundary theory predicts classical bulk evolution are stronger than the criteria usually employed in quantum cosmology. We illustrate this in a homogeneous isotropic minisuperspace model of gravity coupled to a scalar field in which we identify the ensemble of classical histories explicitly. 

\end{abstract}

\maketitle

\tableofcontents

\newpage

\section{Introduction}
 
The dS/CFT correspondence \cite{Balasubramanian2001,Strominger2001,Maldacena2002,Witten2001} conjectures that the wave function of the universe with asymptotic de Sitter (dS) boundary conditions is given in terms of the partition function of a Euclidean CFT deformed by various operators. Explicit realizations of dS/CFT include the duality between Vasiliev gravity in dS and Euclidean $Sp(N)$ vector models \cite{Anninos2011} and, at the semiclassical level, the holographic form of the Hartle-Hawking wave function put forward in \cite{Hertog2011}, building on earlier work \cite{Maldacena2002,Garriga2008,Harlow2011,Maldacena2011, McFadden2009} and further explored in \cite{Castro2012,Banerjee2013,Anninos2012,Anninos2013,Hartle2012a,Hartle2012b}. 

In dS/CFT the arguments of the wave function of the universe are related to external sources in the dual partition function. The dependence of the partition function on the values of these sources thus yields a holographic measure on the space of asymptotically locally de Sitter configurations. But it has remained an open question what is the exact configuration space of deformations on which the holographic wave function in dS/CFT should be defined\footnote{For recent work on this see e.g. \cite{Anninos2012,Anninos2013,Conti2014}.}. Here we study a particular constraint on this configuration space that arises from the condition that the holographic wave function must predict classical Lorentzian space-time evolution in the bulk at least for large spatial volumes.

The classical behavior of geometry and fields on sufficiently large scales follows from the Wheeler-DeWitt (WDW) equation at large volume and thus applies to any wave function satisfying the Hamiltonian constraint. It is usually said that a wave function of the universe predicts classical evolution if its phase varies rapidly compared to its amplitude in all directions in superspace. This criterion is known as the classicality condition and is analogous to the prediction of the classical behavior of a particle in a WKB state in non-relativistic quantum mechanics\footnote{See e.g. \cite{Bousso1998,Hartle2008} for a discussion of this in the context of quantum cosmology.}.

However this derivation of classical evolution does not carry over to the dual partition functions featuring in dS/CFT. This is because the large phase factor of the bulk wave function is absent in the partition function. Specifically it is canceled by the addition of counterterms. This raises the question whether the emergence of classical Lorentzian evolution in the bulk for large spatial volumes can be identified and understood from the dual partition function.

In this paper we derive a sharper set of classicality conditions that are meaningful from an asymptotic viewpoint and which, in particular, can be applied to the Euclidean boundary theory in dS/CFT to verify whether it predicts classical Lorentzian bulk evolution in the large volume limit. 
Our derivation makes use of a canonical transformation to coordinates on superspace that are well-defined from a boundary viewpoint, namely the sources in the dual partition function. After a brief review of the Hartle--Hawking wave function \cite{Hartle1983} and its holographic form \cite{Hertog2011} in Section \ref{hol} we perform this transformation on the wave function in Section \ref{sec:holography} to derive a new wave function that is a function of asymptotically well-defined superspace variables. The relation between the two wave functions resembles and generalizes the Fourier transformation between a wave function in momentum and position space. 

The new wave function no longer contains a phase factor that grows with the spatial three-volume. This is in line with the results of a similar calculation in the context of AdS holography in \cite{Papadimitriou:2010as}, where variables on phase space were identified that make the variational problem well-defined. It was found there that these variables are related to the original AdS fields and momenta by a canonical transformation, and that this transformation is equivalent to holographic renormalization. Specifically, the local boundary terms that regularize the AdS action are exactly the generating function of this canonical transformation. Similarly in the de Sitter context which we consider here, the generating function absorbs or `regularizes' the local phase factor of the wave function.

The expression of the bulk wave function in terms of the sources of the dual enables a sharper and more appealing formulation of dS/CFT in which the wave function of the universe is directly related to the partition functions of deformed Euclidean CFTs. In Section \ref{sec:classicality} we revisit the question of classicality in this context. We analyze the conditions under which the new wave function predicts classical bulk evolution at large spatial volumes and interpret these from a dual viewpoint as the requirement that the vevs must be approximately real. 
This is a stronger condition than the criterion for classical behavior usually employed in quantum cosmology which, in dual terms, involves the sources only. We illustrate this difference in Section \ref{sec:minisuperspace} in a minisuperspace model where we identify the ensemble of classical histories explicitly. 

We perform our calculations in the Hartle-Hawking state because dS/CFT is best developed in this context. However, most of our results apply more generally.

\section{Holographic No-Boundary Measure} \label{hol}

This section reviews the holographic formulation of the no-boundary wave function (NBWF) \cite{Hartle1983} put forward in \cite{Hertog2011}, building on earlier work. The holographic form of the NBWF provides a concrete semiclassical realization of dS/CFT with which we work in the remainder of this paper.

\subsection{No-Boundary Wave Function}

A quantum state of the universe is specified at low energies by a wave function $\Psi$ on the superspace of three-geometries $h_{ij}(\vec{x})$ and matter field configurations $\chi(\vec{x})$ on a closed spacelike three-surface $\Sigma$. The NBWF is given by a sum $\mathcal{C}$ over regular four-geometries $g$ and fields $\phi$ on a four-manifold $M$ with one boundary $\Sigma$, weighted by $\exp(-I_E[g,\phi]/\hbar)$ where $I_E[g,\phi]$ is the Euclidean action. Schematically, we write
\begin{align}
\Psi(h_{ij}, \chi) &= \int_{\cal C} \cD g \cD \phi \exp(-I_E[g(x),\phi(x)]/\hbar)\ . 
\label{eq:psi} 
\end{align}
Here, $g(x)$ (short for $g_{\alpha\beta}(x^\gamma)$) and $\phi(x)$ are the histories of the 4-geometry and matter field. 

To analyze classical bulk evolution from a boundary viewpoint we will work with a toy model consisting of Einstein gravity coupled to a single scalar field, with action
\be
I_E = -\int{\D^4x \sqrt{g}\ \left[ \frac{1}{2\kappa} (R - 2\Lambda) - \frac{1}{2} (\nabla \phi)^2 - V(\phi) \right]} 
+ \frac{1}{\kappa} \int{\D^3x \sqrt{h}\ K} \ , 
\label{eq:I_E} 
\ee
where $\kappa=8\pi G_N$ and $h$, $K$ are the induced metric and extrinsic curvature on $\Sigma$. The integration in \eqref{eq:psi} is carried out along a suitable complex contour which ensures the convergence of \eqref{eq:psi} and the reality of the result. 
In this paper we concentrate on models in which the cosmological constant $\Lambda$ and the potential $V$ in the action \eqref{eq:I_E} are positive. We further assume the potential $V(\phi)$ is quadratic with mass $m^2$ around a minimum at $\phi=0$, where it vanishes.

\subsection{Lorentzian Bulk Evolution}
\label{classens}

In the large three-volume region of superspace, the path integral \eqref{eq:psi} defining the NBWF can be approximated by the method of steepest descent \cite{Hartle2012a}. In this regime the NBWF is approximately given by a sum of terms of the form 
\begin{equation}
\Psi[h_{ij},\chi] \approx  \exp\{(-I_R[h,\chi] +i S[h,\chi])/\hbar\} .
\label{semiclass}
\end{equation}
Here $I_R[h_{ij},\chi]$ and $-S[h_{ij},\chi]$ are the real and imaginary parts of the Euclidean action $I_E[h_{ij},\chi]$ of a saddle point history $(g,\phi)$ on a compact 4-disk $M$ with one boundary $\Sigma$. Metric and field match the real values $(h_{ij},\chi)$ on $\Sigma$ and are otherwise regular on the disk, rendering $(g,\phi)$ generally complex in the interior. 

In the large volume regime boundary configurations $(h_{ij},\chi)$ evolve classically, according to the Lorentzian Einstein equation, because the NBWF oscillates and obeys the classicality conditions that say that its phase $S$ varies rapidly compared to $I_R$ \cite{Hartle2008},
\be
| \vec{\nabla} I_R|\ll |\vec{\nabla} S|    \ .
\label{eq:classicalityIntro}
\ee
This is analogous to the prediction of the classical behavior of a particle in a WKB state in non-relativistic quantum mechanics. 
Thus the NBWF predicts an ensemble of classical, asymptotically de Sitter histories that are the integral curves of $S$ in superspace, with relative probabilities that are proportional to $\exp[-2 I_R(h_{ij},\chi)]$ and conserved under time evolution \cite{Hartle2008}. 

Integral curves are defined by integrating the classical relations relating the momenta $\pi_{(h)ij}(\vec x)$ and $\pi_{(\chi)}({\vec x})$ to derivatives of the action 
\begin{equation}
\label{momenta}
\pi_{(h)ij}(\vec x) =\delta S /\delta h_{ij}(\vec x), \quad \pi_{(\chi)}(\vec x)= \delta S /\delta \chi(\vec x) .
\end{equation}
The momenta are proportional to the time derivatives of $h_{ij}$ and $\chi$. Thus the solutions $h_{ij}({\vec x},t)$ and $\chi({\vec x},t)$ of \eqref{momenta} define field histories ${\hat\phi}(x,t)\equiv\chi(x,t)$ and Lorentzian four-geometries $\hat g_{\alpha\beta}(x,t)$ by
\begin{equation}
\label{lorhist}
ds^2 = -dt^2 + h_{ij}(x,t)dx^i dx^j \equiv {\hat g}_{\alpha\beta}(x,t) dx^\alpha dx^\beta   \ ,
\end{equation} 
in a simple choice of gauge. The real, classical, Lorentzian histories predicted by the NBWF are therefore not the same as the complex saddle points that determine their probabilities. Further, the relations between superspace coordinates and momenta \eqref{momenta} mean that to leading order in $\hbar$, and at any one time, the predicted classical histories do not fill classical phase space. Rather, they lie on a surface within classical phase space of half its dimension. 

\subsection{Holographic No-Boundary Measure} 

By exploiting the complex structure of the no-boundary saddle points one can derive a holographic form of the tree level no-boundary measure \cite{Hertog2011}. To see this we write the saddle point geometries as
 \begin{equation}
\D s^2= (N^2 + N_i N^i) \D \lambda^2 + 2N_i \D x^i \D\lambda + g_{ij} \D x^i \D x^j  \ . 
\label{eucmetric_hij}
\end{equation} 
and introduce the complex time coordinate $d\tau = d\lambda N(\lambda)$. The action \eqref{eq:I_E} of a saddle point history then includes an integral over time $\tau$. Different contours for this time integral give different geometric representations of the saddle point, each giving the same amplitude for the boundary configuration $(h_{ij},\chi)$. This freedom in the choice of contour gives physical meaning to a process of analytic continuation -- not of the Lorentzian histories themselves -- but of the saddle points that define their probabilities.

In \cite{Hertog2011} this freedom of choice of contour was used to identify two different useful representations of the general saddle points corresponding to asymptotically de Sitter universes. In one representation (dS) the interior saddle point geometry behaves as if $\Lambda$ and $V$ were positive and converges towards a real Lorentzian solution that is asymptotically dS. In the other (AdS) the Euclidean part of the interior geometry behaves as if these quantities were negative, and specifies a regular AdS domain wall. Asymptotically Lorentzian de Sitter (dS) universes and Euclidean anti-de Sitter (AdS) spaces are thereby connected in the wave function. This connection can be made explicit using the asymptotic form of the saddle point solutions. If we define the variable $\eta$ by 
\be
\eta(\tau)=i\eta_0 e^{iH\tau} = i\eta_0 e^{-Hy+iHx},
\label{eqn:eta}
\ee
with $H \equiv \sqrt{\Lambda/3}$ and $\eta_0$ an arbitrary scale that we will fix below, then the large volume expansion of the general complex solution of the Einstein equation can be written as
\begin{subequations} 
\label{eqn:asymptoticAnsatz} 
\begin{align}
g_{ij} &= \frac{1}{\eta^2} \left( \gamma_{ij} + \eta^2 \gamma_{(2)ij} + \eta^{3-\sigma}( \gamma_{(-)ij}+\eta \gamma_{(2-)ij}+\ldots) + \eta^3 \gamma_{(3)ij} + \eta^{3+\sigma} \gamma_{(+)ij} + \order{\eta^4} \right)  \ , \label{eqn:asymptoticmetric}    \\
\phi(\eta) &= \eta^{\lambda_-}\gamma^{\lambda_- / 2\sigma} (\alpha +\eta \alpha_{(1)}+\ldots )- \frac{\eta^{\lambda_+}}{\sigma} \gamma^{-\lambda_+ / 2\sigma} (\beta + \eta \beta_{(1)} + \ldots) + \order{\eta^{\lambda_- + 1}}   \ , 
\label{eqn:asymptoticfield}
\end{align} \end{subequations}
where $\lambda_\pm \equiv 3/2 (1 \pm \sqrt{1 - 4m^2 / 9H^2})$, $\sigma \equiv \lambda_+ - \lambda_-$ and $\gamma$ is the determinant of $\gamma_{ij}$.

The asymptotic solutions are specified by the asymptotic equations in terms of the boundary functions $\gamma_{ij}$ and $\alpha$, up to the $\eta^3$ term in \eqref{eqn:asymptoticmetric} and to order $\eta^{\lambda_{+}^{\ }}$ in \eqref{eqn:asymptoticfield}. Beyond this the interior dynamics and the boundary condition of regularity on $\mathcal{M}$ become important.
 
In asymptotically dS saddle points the phases of the fields at the South Pole (SP) -- the center of the 4-disk $\mathcal{M}$ -- are tuned so that $g_{ij}$ and $\phi$ become real for small $\eta$ along a vertical line $x= x_{TP}$ in the complex $\tau$-plane. Equations \eqref{eqn:asymptoticAnsatz} show that along this curve the complex saddle point tends to a real, asymptotically dS history. However since the expansions are analytic functions of $\eta$ there is an alternative asymptotically vertical contour located at $x_A=x_{TP}-\pi/2H$ along which the metric $g_{ij}$ is also real, but with the opposite signature. Along this contour the saddle point geometry \eqref{eqn:asymptoticmetric} is asymptotically Euclidean AdS. Hence a contour which first runs along the $x=x_A$ line and then cuts horizontally to the endpoint $\tau = \upsilon$ provides a representation of the saddle points in which their interior geometry consists of a regular Euclidean AdS domain wall that makes a smooth transition to an asymptotically dS universe.

\begin{figure}[t]
\begin{center}
\begin{tikzpicture}
    \draw[->, thick] (-35pt,0pt) -- (120pt,0pt) node[below] {x};
    \draw (0, -5pt) node[below] {SP};
    \draw[->, thick] (0pt, -5pt) -- (0pt,180pt) node[left] {y};
    \draw[->-=.5, ultra thick] (-.8pt,0pt) -- (90.8pt,0pt) node[below] {$x_{TP}$};
    \draw[->-=.5, ultra thick] (90pt,-.8pt) to (90pt,160.8pt);
    \draw[-, ultra thick, dotted] (0pt,-.8pt) to[out=90,in=-60] (-15pt,60pt);
    \draw[->-=.5, ultra thick] (-15pt,60pt) to[out=120,in=-90] (-30pt,120pt);
    \draw[-, ultra thick] (-30pt,120pt) -- (-30pt,160.8pt);
    \draw[-, dashed] (-30pt,0pt) -- (-30pt,160.8pt);
    \draw (-30pt,0) node[below] {$x_A$};
    \draw[->-=.5, ultra thick] (-30.8pt,160pt) to (90.8pt,160pt);
    \draw[solid,fill] (90pt,160pt) circle (2pt) node[right] {$\upsilon$};
    \draw (90pt, 120pt) node[right] {dS};
    \draw (-30pt, 120pt) node[left] {AdS};
    \draw (120pt, 170pt) -- (110pt, 170pt) -- (110pt, 185pt) node[below right] {$\tau$};
\end{tikzpicture}
\end{center}
\caption{Two representations in the complex $\tau$-plane of the same no-boundary saddle point associated with an inflationary universe. Along the vertical part of the AdS contour the geometry is an asymptotically AdS, spherically symmetric domain wall with a complex scalar field profile. Along the vertical branch of the dS contour the saddle point tends to a Lorentzian, inflationary universe. The logarithm of the amplitude of this universe is given by the AdS domain wall action. The horizontal branch of the AdS contour connecting AdS to dS automatically regularizes the AdS action.}
\label{contour}
\end{figure}
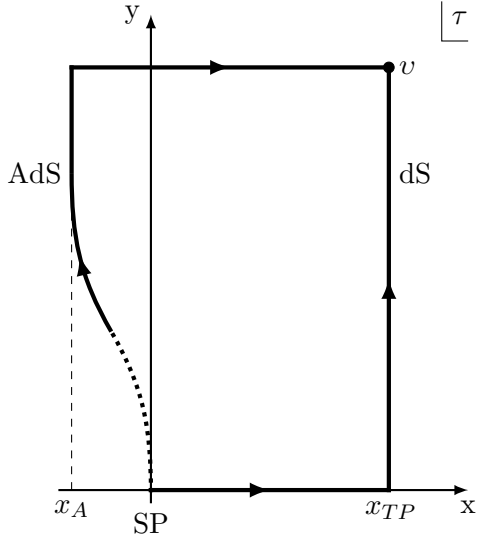

Figure \ref{contour} illustrates this for an $O(4)$ invariant saddle point in which the scalar field $\phi(0)$ at the SP at $\tau=0$ is significantly displaced from the minimum of its potential. The symmetry allows one to identify both contours explicitly (as opposed to only asymptotically for general saddle points). The dS contour first runs along the real axis to a turning point at $x_{TP} = \pi/2V(\phi(0))$ and then vertically to the endpoint $\upsilon$. This corresponds to the usual saddle point representation where a deformed four-sphere is smoothly joined onto an approximately real, inflationary universe in which the scalar slowly rolls down to its final value.

The AdS contour starts vertically but gradually moves away from the dS contour to $x= x_A$ at large $y$. Along this part of the contour the saddle point is an asymptotically AdS, spherically symmetric domain wall with a complex scalar field profile in the radial direction $y$. The complex transition region along the horizontal branch of the contour smoothly interpolates between the AdS and the dS domain. This contour has the same endpoint $\upsilon$, the same action, and makes the same predictions, but the saddle point geometry is different.

The real part of the Euclidean action along $x=x_A$ has the usual AdS divergences for large $y$. However, along the $x=x_{TP}$ curve the real part of the action is asymptotically constant and hence does not grow parametrically with $y$. Therefore, the horizontal branch of the AdS contour regulates the divergences of the AdS action. 

It can be shown \cite{Hertog2011} that for general saddle points the divergent terms in the action of the horizontal part are precisely the regulating counterterms plus a universal phase factor $S_{ct}$. Moreover the action integral along the horizontal branch of the AdS contour does not contribute to the amplitude in the large $y$ limit \cite{Hertog2011}. This means that the probabilities for all Lorentzian asymptotically dS histories in the NBWF are fully specified by the regularized action of the interior asymptotic AdS regime of the saddle points. Specifically,
\be
I_E [\eta(\upsilon), h_{ij}, \chi] = -I^{\rm reg}_{aAdS} [\bar{\gamma}_{ij}, \bar{\alpha}] + iS_{ct}[\eta(\upsilon), h_{ij}, \alpha] + \order{\eta(\upsilon)} \ . \label{eqn:holographicAction}
\ee
Here $I^{\rm reg}_{aAdS}$ is the $y \rightarrow \infty$ limit of the regulated asymptotic AdS action. The barred quantities $\bar{\alpha}$ and $\bar{\gamma}_{ij}$ are the coefficients in the Fefferman-Graham expansion of the saddle points along the asymptotic AdS branch of the contour, in terms of the radial AdS coordinate $z = -i\eta$. For example
\begin{align}
\phi(z) =& \bar{\alpha} \bar{\gamma}^{\lambda_- / 2\sigma} z^{\lambda_-} - \frac{\bar{\beta}}{\sigma} \bar{\gamma}^{-\lambda_+ / 2\sigma} z^{\lambda_+}	\ .
\end{align}
Hence $\bar{\alpha} \equiv \alpha e^{-i \pi \lambda_+ \lambda_- / \sigma}$ and $\bar{\gamma}_{ij} \equiv -\gamma_{ij}$. The minus sign in front of $I^{\rm reg}_{aAdS}$ in \eqref{eqn:holographicAction} is connected to the fact that the NBWF behaves as a decaying wave function along the AdS branch of the contour \cite{Gabriele:2015gca}. Euclidean AdS/CFT relates this term to the partition function of a dual field theory. This yields the following holographic form of the semiclassical NBWF in the large volume limit,
\be
\Psi[h_{ij}, \chi]= Z^{-1}_{QFT}[\bar{\gamma}_{ij},\bar{\alpha}] \exp(iS_{ct}[h_{ij}, \chi]/\hbar)   \ .
\label{eq:dSCFT}
\ee

\subsection{Classicality from a holographic viewpoint?}

Equation \eqref{eq:dSCFT} shows that to leading order in $\hbar$ the probabilities of all asymptotically locally dS histories in the no-boundary state are given by the inverse of the partition function of a Euclidean AdS/CFT dual field theory defined on the boundary surface $\Sigma$. The arguments $(h_{ij}, \chi)$ of the wave function enter as external sources $(\bar{\gamma}_{ij}, \bar{\alpha})$ in $Z_{QFT}$. The dependence of the partition function on the values of the sources therefore gives a holographic measure on asymptotically locally de Sitter configurations. 

The probability of each individual history is conserved under scale factor evolution as a consequence of the Wheeler-DeWitt equation. However, the wave function itself oscillates rapidly in the large volume regime. In fact it appears that the large phase factor in \eqref{eq:dSCFT} is crucial in order for the classicality conditions \eqref{eq:classicalityIntro} to hold and hence for the wave function to predict classical Lorentzian evolution in the first place. This raises the question whether the emergence of classical space-time evolution can be understood from the dual partition function.

To answer this we will in the next section rewrite the asymptotic wave function in terms of superspace coordinates $(\bar{\gamma}_{ij}, \bar{\alpha})$ that are natural and meaningful from an asymptotic viewpoint and in particular enter as sources in $Z$. Since this involves a map between two symplectic manifolds we will work in the Hamiltonian formulation of the theory, where the symplectic structure is manifest. The relation between the two wave functions resembles and generalizes the Fourier transformation between a quantum mechanical wave function in momentum and position space. In Section \ref{sec:classicality} we then revisit the classicality conditions in terms of the new superspace coordinates. 

We note that a calculation in the same spirit was done in the context of AdS holography in \cite{Papadimitriou:2010as}, where variables on phase space were identified that make the variational problem well-defined. These variables were found to be related by a canonical transformation to the original AdS fields and momenta. It was established that this transformation is equivalent to holographic renormalization. Furthermore, the generating function of this canonical transformation coincides exactly with the local boundary terms that regularize the AdS action. In the next section we show that this conclusion carries over to the dS case where the local boundary terms appear to play a physical role as discussed above.

\section{Asymptotic Wave Function} \label{sec:holography}

We now derive the NBWF as a function of a new set of variables that remain finite and physically meaningful in the large volume limit. These variables come in canonically conjugate pairs so they are related to the original fields and momenta through a canonical transformation. 

\subsection{The Hamiltonian NBWF} 
We work with the ADM form of the saddle point metrics \eqref{eucmetric_hij} and with the NBWF in Hamiltonian form\footnote{To simplify notation we set the Hubble constant $H=1$ in this section.}, 
\begin{align}
\Psi(h_{ij}, \chi) & \propto  \int  {\cD g_{ij} \cD \phi \cD \pi_{(g)}^{ij} \cD \pi_{(\phi)} \cD N \cD N^i \ e^{\frac{1}{\hbar} \int{\D\lambda \int{\D^3x\ \left[ i \pi_{(\phi)} \dot{\phi} + i \pi_{(g)}^{ij} \dot{g}_{ij} - \sqrt{g}\left( N H + N^i H_i \right) \right] }} } } \ ,
\label{eqn:psi_H}
\end{align}
where $\pi_{(\phi)}$ and $\pi_{(g)}^{ij}$ are the conjugate momenta of the scalar field and the metric, a dot means a derivative with respect to $\lambda$ and where we introduced
\begin{align}
H &=  \frac{2\kappa}{g} \left( \pi_{(g)}^{ij} \pi_{(g)ij} - \frac{\tr(\pi_{(g)})^2}{2} \right) + \frac{(\pi_{(\phi)})^2}{2g} + \frac{1}{2\kappa} \left( -{ }^{(3)}R + 2\Lambda \right) + \frac{1}{2}  g^{ij} \partial_i \phi \partial_j \phi + V(\phi)  \ , \nonumber \\
H_i &= -2i \cD^j \left( \frac{\pi_{(g)ij}}{\sqrt{g}} \right) + i \frac{\pi_{(g)i}^{j}}{\sqrt{g}} \partial_j \phi \ ,
\end{align}
with $\cD^j$ the covariant derivative on slices of constant $\lambda$. Also, ${ }^{(3)}R$ is the three dimensional Ricci scalar constructed from $g_{ij}$. Performing the Gaussian integrations over $\pi_{(\phi)}$ and $\pi_{(g)}^{ij}$ by substituting their extremizing values yields the original action \eqref{eq:I_E}.

Variation of the action with respect to $N$ and $N_i$ yields the first-class Hamiltonian and momentum constraints. One can use the gauge freedom of coordinate reparametrizations to fix the values of these fields, e.g. $N^i=0$ and $N=1$. Concerning the lapse, a change $N(\lambda) \to N(\lambda) + f'(\lambda)$ can be absorbed in a redefinition of the time coordinate $\lambda$ to keep the metric invariant. However, the constraint that the range of $\lambda$ remains unchanged means $\upsilon \equiv \int{N \D \lambda}$ is invariant under gauge transformations. To differentiate between the physical data $\upsilon$ and all other -- gauge dependent -- information in $N$, we introduce the following variables
\begin{align}
\tau(\lambda) &= \int_0^\lambda{\D \lambda'\ N(\lambda)}    \ , &   \int{\cD N(\lambda)} \rightarrow \int{\cD \tau(\lambda)}    \ .
\end{align}
The only physical information contained in $\tau(\lambda)$ is $\tau(1) = \upsilon$. Therefore, the path integral over $N$ reduces to a physically irrelevant constant of proportionality -- the volume of the gauge group -- and an ordinary integral $\int{\D \upsilon}$ over all possible values of $\int{\D \lambda\ N(\lambda)}$.\footnote{This can also be seen from an explicit discretization of the measure. Writing
\begin{align}
\int \cD N = \lim_{J \to \infty} \int{ \left( \prod_{m=0}^{J} \D N_m \right)} \ ,
\end{align}
where the subscript $m$ labels the $\lambda$-slice, one can consider the following change of variables:
\begin{align}
M_m &\equiv \sum_{n = 0}^m N_n  \ , &   \prod_{m=0}^J \D N_m &= \prod_{m=0}^J \D M_m    \ . \label{eq:NtoM}
\end{align}
The second equation follows from the fact that the Jacobian is 1, since the transformation matrix is triangular with only 1's on the diagonal. One can now easily separate the physical quantity $\upsilon = M_J$ from the rest. Hence for $m < J$, $M_m$ has no physical significance and we can gauge it away. Therefore,
\begin{align}
\int \cD N = \lim_{J \to \infty} \int{\left( \prod_{m=0}^{J-1} \D M_m \right) \D M_J} =  \lim_{J \to \infty} \int{\left( \prod_{m=0}^{J-1} \D M_m \right) \D \upsilon} \propto \int{\D \upsilon} \ .
\end{align}
}

The asymptotic momenta in the Hamiltonian version of the action can be expressed in terms of the coefficients in the Fefferman-Graham (FG) expansion \eqref{eqn:asymptoticAnsatz} as follows:
\begin{align}
\pi_{(\phi)} &= i \sqrt{g} \left( \phi' - N^i \partial_i \phi \right) = -\lambda_- \alpha \gamma^{\lambda_+/2\sigma} \eta^{-\lambda_+} + \lambda_+ \frac{\beta}{\sigma} \gamma^{-\lambda_-/2\sigma} \eta^{-\lambda_-} + \order{\eta^{-\lambda_+ + 1}}  \ , \label{eq:phiMomentum} \\
{{\pi_{(g)}}^i}_j &= \pi_{(g)}^{ik} g_{kj} = i \frac{\sqrt{g}}{2\kappa} \left( K^{ik} - K g^{ik} \right) g_{kj}  \nonumber \\
&= \frac{\sqrt{g}}{2\kappa} \left[ -2 \delta^i_j - \eta^2 \left( {\gamma_{(2)}}^i_j - \gamma_{(2)} \delta^i_j \right) - \lambda_- \eta^{3-\sigma} \left( {\gamma_{(-)}}^i_j - 			\gamma_{(-)} \delta^i_j \right) \right. \nonumber \\
&\qquad\qquad\quad \left. - \frac{3}{2} \eta^3 \left( {\gamma_{(3)}}^i_j - \gamma_{(3)} \delta^i_j \right) - \lambda_+ \eta^{3+\sigma} \left( {\gamma_{(+)}}^i_j - \gamma_{(+)} 			\delta^i_j \right) + \order{\eta^4} \right]  \ , \label{eq:metricMomentum}
\end{align}
where prime denotes a derivative with respect to $\tau$. These relations follow from the on-shell expression of the momenta in terms of time derivatives of the fields and will be useful below to motivate the change of coordinates we apply there.

Note that the equations of motion of the momenta provide an alternative way to determine the coefficients in the FG expansion order by order, with $\gamma_{ij}$, $\gamma_{(3)ij}$, $\alpha$ and $\beta$ as the only independent coefficients \cite{Starobinsky:1982mr}. For instance, the equation for $\pi_{(\phi)}$ implies $\lambda_\pm = 3/2 \left( 1 \pm \sqrt{1 - 4 m^2 / 9} \right)$. Similarly, the leading-order equation for $\pi_{(g)}^{ij}$ is satisfied if $\Lambda = 3$ (remember that we are working in units of the de Sitter length) and the next orders imply
\begin{align}
\gamma_{(-)ij} &= -\frac{\kappa}{4} \gamma^{\lambda_-/\sigma} \alpha^2 \gamma_{ij}    \ , &   \gamma_{(2)ij} &= \left( R_{(\gamma)ij} - \frac{1}{4} R_{(\gamma)} \gamma_{ij} \right)   \ , &   \gamma_{(+)ij} &= -\frac{\kappa}{4} \gamma^{-\lambda_+/\sigma} \frac{\beta^2}{\sigma^2} \gamma_{ij} \ , \label{eq:EoMSol}
\end{align}
where $R_{(\gamma)ij}$ and $R_{(\gamma)}$ are the Ricci tensor and scalar constructed from $\gamma_{ij}$. Finally, defining $\pi_{(\gamma)}^{ij}$ as the coefficient of the ${\cal O}(\eta^3)$-term in \eqref{eq:metricMomentum}, i.e. $\pi_{(\gamma)}^{ij} \equiv \frac{3\sqrt{\gamma}}{4\kappa} (\gamma_{(3)} \gamma^{ij} - \gamma_{(3)}^{ij})$, we note that the Hamiltonian and momentum constraints require that 
\begin{align}
\tr\ \pi_{(\gamma)} &= \frac{\lambda_+\lambda_- \alpha \beta}{\sigma}   \ , &   \cD_{(0)}^j \pi_{(\gamma) ij} &= \frac1{2\sigma} (\lambda_- \alpha \partial_i\beta+\lambda_+\beta \partial_i\alpha)   \ .
\end{align}

\subsection{Canonical transformation to asymptotic coordinates}

We now proceed by describing the canonical transformation to variables that are meaningful asymptotically. Inspired by the FG expansions \eqref{eqn:asymptoticAnsatz}, we introduce a new set of coordinates and conjugate momenta $(\Gamma_{ij}, \Pi_{(\Gamma)}^{ij}, A, B,\eta)$ on extended phase space as follows:
\begin{align}
g_{ij} &\equiv \frac1{\eta^2} \left[ \Gamma_{ij} + \eta^2 \left( R_{(\Gamma)ij} - \frac14 R_{(\Gamma)} \Gamma_{ij} \right) - \frac{\kappa}4 \eta^{3-\sigma} A^2 \Gamma^{\lambda_-/\sigma} \Gamma_{ij} \right.	\nonumber \\
&\qquad\quad \left.  + \frac{2\kappa}{3\sqrt{\Gamma}} \eta^3 (\Pi_{(\Gamma)} \Gamma_{ij} - 2\Pi_{(\Gamma) ij}) - \frac{\kappa}{4\sigma^2} \eta^{3+\sigma} B^2 \Gamma^{-\lambda_+/\sigma} \right]	\ ,    \nonumber \displaybreak[0] \\
\pi_{(g)}^{ij} &\equiv \frac1{2\kappa} \frac{\sqrt{\Gamma}}{\eta} \left[ -2 \Gamma^{ij} + \eta^2 \left( R_{(\Gamma)}^{ij} - \frac14 R_{(\Gamma)} \Gamma^{ij} \right) \right] + \frac{\sigma - 2}8 \eta^{2-\sigma} A^2 \Gamma^{3/2\sigma} \Gamma^{ij} \nonumber \\
&\quad + \frac{\eta^2}3 \left( \Pi_{(\Gamma)} \Gamma^{ij} - \Pi_{(\Gamma)}^{ij} \right) - \frac{\sigma + 2}{8\sigma^2} \eta^{2+\sigma} B^2 \Gamma^{-3/2\sigma} \Gamma^{ij}\ , \nonumber
\end{align} \vspace{-12pt} 
\begin{align}
\phi &\equiv A \Gamma^{\lambda_- / 2\sigma} \eta^{\lambda_-} - \frac{B}{\sigma} \Gamma^{-\lambda_+ / 2\sigma} \eta^{\lambda_+}   \ , &   \pi_{(\phi)} &\equiv -\lambda_- A \Gamma^{\lambda_+ / 2\sigma} \eta^{-\lambda_+} + \frac{\lambda_+}{\sigma} B \Gamma^{-\lambda_- / 2\sigma} \eta^{-\lambda_-}  \ , \label{eq:asymptoticVars}
\end{align}
where $R_{(\Gamma)}$ is the Ricci scalar associated with the metric $\Gamma_{ij}$ and $\Gamma$ is its determinant. We raise and lower indices of the new phase space coordinates by acting with $\Gamma_{ij}$ and its inverse $\Gamma^{ij}$.

Note that $(\Gamma_{ij}, \Pi_{(\Gamma)}^{ij}, A, B)$ are functions of the scale factor variable $\eta$, but $\Gamma_{ij} \to \gamma_{ij}$, $\Pi_{(\Gamma)}^{ij} \to \pi_{(\gamma)}^{ij}$, $A \to \alpha$ and $B \to \beta$ when $\eta \to 0$. The $\eta$-dependence merely incorporates the higher order terms in the FG expansions\footnote{The $\eta$-dependence of the new variables guarantees that \eqref{eq:asymptoticVars} holds exactly at all times. Specifically, $A$ contains all terms that are higher order in $\eta$, such as the terms with coefficient $\alpha_1$ in \eqref{eqn:asymptoticAnsatz} and higher order terms of the form $\eta^{\lambda_- + n}$ with $n>0$. Similarly the corrections to B are of the form $\eta^{\lambda_{+}+n}$. The same goes for $\Gamma_{ij}$.}.

Since we only consider compact three-geometries, we can fix the scale of $\eta$ in \eqref{eqn:eta} by choosing $\eta_0=1/\sqrt{\textrm{Vol}(h_{ij})}$. This guarantees that $\Gamma_{ij}$ has volume 1 at late times.

What we would like to find, is a generating function $f$, such that 
\begin{align}
\pi_{(g)}^{ij} \D g_{ij} + \pi_{(\phi)} \D\phi + \sqrt{g} H \frac{\D\eta}{\eta} &= \Pi_{(\Gamma)}^{ij} \D\Gamma_{ij} + B \D A + \sqrt{\Gamma} \tilde{H} \frac{\D\eta}{\eta} + \D f + \mathcal{P}(\eta) \ , \label{eqn:cantransf}
\end{align}
where $\mathcal{P}(\eta)$ is a function consisting of $dA$, $dB$, $d\Gamma$ and $d\Pi_{(\Gamma)}$ terms that cannot be written in the form of one of any of the other terms in \eqref{eqn:cantransf}. We will find that $\mathcal{P}(\eta)$ is higher order in $\eta$, as expected. Using the definition of the dynamical asymptotic variables \eqref{eq:asymptoticVars}, the left-hand side of \eqref{eqn:cantransf} can be expressed in terms of $\Gamma_{ij}$, $\Pi_{(\Gamma)}^{ij}$, $A$ and $B$. For the scalar field variables we get
\begin{align}
\pi_{(\phi)} \D\phi &= B \D A - \D\left( \lambda_- \Gamma^{3/2\sigma} \eta^{-\sigma} \frac{ A^2}{2} - \frac{\lambda_-}{\sigma} A B + \frac{\lambda_+}{\sigma^2} \Gamma^{-3/2\sigma} \eta^\sigma \frac{B^2}{2} \right)    \nonumber \\
&\quad + \left( \frac{\lambda_-}{4} \eta^{-\sigma} A^2 \Gamma^{3/2\sigma} + \frac{\lambda_+}{4\sigma^2} \eta^\sigma B^2 \Gamma^{-3/2\sigma} \right) \Gamma^{ij} \D \Gamma_{ij} \nonumber \\
&\quad - \left( \frac{3\lambda_-}{2} A^2 \Gamma^{3/2\sigma} \eta^{-\sigma} - 2 \frac{\lambda_+ \lambda_-}{\sigma} AB + \frac{3\lambda_+}{2} \frac{B^2}{\sigma^2} \Gamma^{-3/2\sigma} \eta^\sigma \right) \frac{\D \eta}{\eta}   \ ,
\end{align}
where we have used that $\sqrt{g}$ can be expanded as $\sqrt{\Gamma}/\eta^3$ and terms of higher order in $\eta$,
\begin{align}
\sqrt{g} &= \frac{\sqrt{\Gamma}}{\eta^3} \left( 1 + \frac{\eta^2 R_{(\Gamma)} }{8}- \frac{\eta^{3-\sigma} 3 \kappa \Gamma^{\lambda_-/\sigma} A^2 }{2} + \frac{\eta^3 \kappa \Pi_{(\Gamma)}}{3\sqrt{\Gamma}} - \frac{\eta^{3+\sigma} 3\kappa \Gamma^{-\lambda_+/\sigma}}{8\sigma^2} + \order{\eta^4} \right)    \label{eq:sqrtg}   \ .
\end{align}
Similarly, the gravitational part of the symplectic form is
\begin{align}
\pi_{(g)}^{ij} \D g_{ij} &= -\frac{2}{\kappa} \D\sqrt{g} + \left[ \frac1{2\kappa} \frac{\sqrt{\Gamma}}{\eta} \left( \frac12 R_{(\Gamma)} \Gamma^{ij} - R_{(\Gamma)}^{ij} \right) \right.    \nonumber \\
&\quad \left. - \frac{\lambda_-}{4} \eta^{-\sigma} A^2 \Gamma^{3/2\sigma} \Gamma^{ij} + \Pi_{(\Gamma)}^{ij} - \frac{\lambda_+}{4\sigma^2} \eta^{\sigma} B^2 \Gamma^{-3/2\sigma} \Gamma^{ij} \right] \D \Gamma_{ij}  \\
&\quad + \left( -\frac1{2\kappa} \frac{\sqrt{\Gamma}}{\eta} R_{(\Gamma)} + \frac{3\lambda_-}2 \eta^{-\sigma} A^2 \Gamma^{3/2\sigma} - 2 \Pi_{(\Gamma)} + \frac{3\lambda_+}{2\sigma^2} \eta^{\sigma} B^2 \Gamma^{-3/2\sigma} \right) \frac{\D\eta}{\eta} + \mathcal{P}(\eta)  \ ,    \nonumber \label{eq:pidg}
\end{align}
where $\sqrt{g}$ is given in \eqref{eq:sqrtg}. Adding these expressions one sees that the terms proportional to $A^2 \D \Gamma_{ij}$ and $B^2 \D \Gamma_{ij}$ cancel. Furthermore, the term proportional to $d\Gamma_{ij}/\eta$ in \eqref{eq:pidg} is a total derivative,
\begin{align}
\frac1{2\kappa} \frac{\sqrt{\Gamma}}{\eta} \left[ \left( \frac12 R_{(\Gamma)} \Gamma^{ij} - R_{(\Gamma)}^{ij} \right) \D \Gamma_{ij} - R_{(\Gamma)} \frac{\D\eta}{\eta} \right] &= \D \left( \frac1{2\kappa} \frac{\sqrt{\Gamma}}{\eta} R_{(\Gamma)} \right)    \ .
\end{align}
Finally, expanding $\sqrt{g}$ in the first term of \eqref{eq:pidg} using \eqref{eq:sqrtg}, one obtains \eqref{eqn:cantransf} with
\begin{align}
f &= \frac{\sqrt{\Gamma}}{\kappa} \left( \frac{-2}{\eta^3} + \frac{1}{4\eta} R_{(\Gamma)} \right) + \frac{\sigma}{4} \eta^{-\sigma} \Gamma^{3/2\sigma} A^2 - \frac23 \Pi_{(\Gamma)} + \frac{\lambda_-}{\sigma} A B - \frac{1}{4\sigma} \Gamma^{-3/2\sigma} \eta^\sigma B^2  \label{eq:generatingFunction}
\end{align}
and
\begin{align}
\sqrt{\Gamma} \tilde{H} \frac{\D\eta}{\eta} &= \sqrt{g} H \frac{\D\eta}{\eta} + 2 \left( \frac{\lambda_+ \lambda_-}{\sigma} AB - \Pi_{(\Gamma)} \right) \frac{\D\eta}{\eta}	\ . \label{eq:KHamiltonian}
\end{align}
Furthermore, $\mathcal{P}(\eta)$ indeed only contains terms that are higher order in $\eta$. Note that the additional term in the new Hamiltonian $\tilde{H}$ vanishes exactly when the Hamiltonian constraint is applied. This means that the physical constraint remains the same, as it should.

\subsection{A new wave function \texorpdfstring{$\tilde \Psi$}{\tilde \Psi}}

We now implement the asymptotically canonical transformation to the variables $(\Gamma_{ij}, \Pi_{(\Gamma)}^{ij}, A, B,\eta)$ at the level of the wave function and write the NBWF in terms of these new variables. In the $\eta \to 0$ limit, this reduces to an asymptotic wave function $\tilde{\Psi}_{as}$ which is a function of $(\gamma_{ij},\alpha)$\footnote{The connection between wave functions in terms of different, canonically related variables is described in more detail and more generally in Appendix \ref{sec:QMtransfo}.}.

The exponent in the Hamiltonian version \eqref{eqn:psi_H} of the path-integral has the same structure as the symplectic one-form in \eqref{eqn:cantransf}, with differentials replaced by time derivatives, and without the $\mathcal{P}(\eta)$ term in \eqref{eqn:cantransf}. Including this as a correction to the new Hamiltonian $\tilde{H}$ which vanishes in the $\eta \to 0$ limit, we can use \eqref{eqn:cantransf} to rewrite the exponent in \eqref{eqn:psi_H} in terms of the new variables. Furthermore, since the Jacobian of a canonical transformation equals one, the measure is left invariant\footnote{This again only holds up to $\order{\eta}$. We will not consider those correction terms because they higher order in $\hbar$.}. Hence we can write, for $\eta_* = \eta(\upsilon)$,
\begin{align}
\Psi(h_{ij}, \chi) \propto& \int \limits_{\mathcal{C}}\cD \Gamma_{ij} \cD \Pi_{(\Gamma)}^{ij} \cD A \cD B \D \log(\eta_*) \ e^{ \frac{i}{\hbar}\int \D \eta \int{\D^3 x\ \frac{\D f}{\D\eta}(\Gamma_{ij}, \Pi_{(\Gamma)}^{ij} A, B,\eta)} } \nonumber \\
&\qquad \cdot e^{+ \frac{i}{\hbar} \int{ \D\eta \int{\D^3x\ \left( B \frac{\D A}{\D \eta} + \Pi_{(\Gamma)}^{ij} \frac{\D \Gamma_{ij}}{\D \eta} + \sqrt{\Gamma} \frac{\tilde{H}}{\eta} \right) } } } \ ,
\end{align}
where the integral is still over the class $\mathcal{C}$ of histories that obey the no-boundary conditions of regularity and compactness and that match $(h_{ij}, \chi)$ on the boundary $\Sigma$. That is,
\be\label{bcs}
g_{ij}(\eta_*, \Gamma_{ij}, \Pi_{(\Gamma)}^{ij}, A, B) = h_{ij}, \qquad \phi(\eta_*, \Gamma_{ij}, \Pi_{(\Gamma)}^{ij}, A, B) = \chi \ .
\ee
Solving the conditions \eqref{bcs} yields an expression of the boundary values of the momenta in terms of the old coordinates, the new coordinates and $\eta_*$,
\begin{align}\label{bcsm}
\Pi_{(\Gamma)}^{ij} (\Gamma_{ij},A,h_{ij}, \chi,\eta_*) \ , \qquad B(\Gamma_{ij},A,h_{ij}, \chi,\eta_*) \ .
\end{align} 
The term with the generating function is a boundary term at $\eta = \eta_*$. Substituting the solutions \eqref{bcsm} in $f$ defines a new function $\tilde{f}$ that does not depend on the momenta but is a function of the coordinates only -- both the original ones and the new asymptotic coordinates, 
\begin{align}
\tilde f &= \frac{\sqrt{\Gamma}}{\kappa \eta_*^3} \left[ 1 + \eta_*^2 \left( \Gamma^{ij} h_{ij} + \frac{R_{(\Gamma)}}2 \right) \right] - \frac\sigma2 \eta_*^{-\sigma} A^2 \Gamma^{3/2\sigma} + \sigma \eta_*^{-\lambda_+} A \chi \Gamma^{\lambda_+ / 2\sigma} - \frac{\lambda_+}2 \frac{\sqrt{\Gamma}}{\eta_*^3} \chi^2  \ . \label{eq:tildeF}
\end{align}
Finally, by inserting the identities
\begin{align}
\int \cD A &= \int \D A_* \int_{A(\eta_*)=A_*} \cD A  \ , &   \int \cD \Gamma_{ij} &= \int \D\Gamma_{*ij} \int_{\Gamma_{ij}(\eta_*)=\Gamma_{*ij}} \cD \Gamma_{ij}    \ ,
\end{align}
in the wave function we can take the generating function $\tilde f$ outside the path integral, because it depends only on the (fixed) boundary values. This yields
\begin{align}
\Psi(h_{ij}, \chi) = \int{\D \Gamma_{*ij} \D A_* \D \log(\eta_*)\ e^{\frac{i}{\hbar} \int{\D^3 x\ \tilde{f}(\Gamma_{*ij}, A_*, h_{ij}, \chi, \eta_*)}}\ \tilde{\Psi}(\Gamma_{*ij}, A_*, \eta_*)} \ ,   \label{eqn:transformedNBWF}
\end{align}
with the no-boundary wave function $\tilde{\Psi}(\Gamma_{*ij},A_*,\eta_*)$ in terms of the asymptotic variables given by
\begin{align}
\tilde{\Psi}(\Gamma_{*ij}, A_*, \eta_*) &\equiv \int_{\mathcal{C}} {\cD \Gamma_{ij} \cD \Pi_{(\Gamma)}^{kl} \cD A \cD B \ e^{ \frac{i}{\hbar}\int{ \D\eta \int{\D^3x\ \left(  B \frac{\D A}{\D \eta} + \Pi_{(\Gamma)}^{ij} \frac{\D \Gamma_{ij}}{\D \eta} + \sqrt{\Gamma} \frac{\tilde{H}}{\eta} \right) } } } } \ .   \label{eq:tildePsi}
\end{align}
Since the transformation to new coordinates is canonical for $\eta \to 0$, the structure of the path integral representing the new wave function $\tilde{\Psi}$ is similar to that of the original wave function, only with the new Hamiltonian $\tilde H$ replacing the original $H$.

The relation \eqref{eqn:transformedNBWF} can be inverted by considering \eqref{eqn:cantransf} in the other direction, with the generating function $f$ subtracted from both sides. A similar derivation, starting by inverting the relations \eqref{eq:asymptoticVars}, then yields
\begin{align}
\tilde{\Psi}(\Gamma_{*ij}, A_*, \eta_*) &= \int{\D h_{ij}\D \chi\ e^{-\frac{i}{\hbar} \int{\D^3 x\ \tilde{f}(\Gamma_{*ij}, A_*, h_{ij}, \chi, \eta_*)}} \Psi(h_{ij}, \chi)}   \ .
\end{align}
The derivation of \eqref{eqn:transformedNBWF} together with the definition of $\tilde \Psi$ is the central result of this paper. The new wave function is obtained from the original wave function by a transformation that generalizes the Fourier transform. The generating function $f$ is a finite polynomial, given in \eqref{eq:generatingFunction}. Furthermore, since the dynamical $\eta$-dependent asymptotic variables tend to constants at late times, $\tilde \Psi$ converges to the asymptotic wave function $\tilde{\Psi}_{as}(\gamma_{ij}, \alpha)$ where $(\gamma_{ij},\alpha)$ do not depend on $\eta$. This is the form of the wave function of the universe that is directly related to and potentially computed with dS/CFT techniques.

To gain further intuition about the relation between the two formulations of the wave function we consider the semi-classical approximation of both $\Psi(h_{ij}, \chi) $ and $\tilde{ \Psi}(\Gamma_{*ij}, A_*,\eta_*)$. In the semi-classical approximation $\Psi(h_{ij}, \chi)$ can be written as
\begin{align}
\Psi(h_{ij}, \chi) &\propto e^{ -\frac1\hbar I_{\textrm{extr}}[h_{ij},\chi] } \ ,
\end{align}
where $I_{\textrm{extr}}$ is the action of a regular compact saddle point, i.e. an extremizing solution to the equations of motion, that satisfies the boundary condition that $g_{ij} = h_{ij}$ and $\phi = \chi$ at the boundary $\Sigma$ where $\eta =\eta_*$. For simplicity we assume here there is only one such saddle point. 

Similarly, we substitute the semi-classical form of $\tilde{ \Psi}(\Gamma_{*ij}, A_*, \eta_*)$ in \eqref{eqn:transformedNBWF},
\begin{align}
\int{\D\Gamma_{*ij} \D A_* \D \log(\eta_*)\ e^{ \frac i\hbar \int{ \D^3x\ \tilde{f}(\Gamma_{*ij}, A_*, h_{ij}, \chi, \eta_*)}} e^{ -\frac1\hbar \tilde{I}_{\textrm{extr}}[\Gamma_{*ij}, A_*, \eta_*] } }  \ . \label{eqn:steepest1}
\end{align}
Solving the remaining integral in the steepest descent approximation yields the following relations,
\begin{align}
i \int d^3x \frac{\partial \tilde{f}}{\partial \Gamma_{*ij}} &= \frac{\partial \tilde{I}}{\partial \Gamma_{*ij}}\ , &   i \int d^3x \frac{\partial \tilde{f}}{\partial A_*} &= \frac{\partial \tilde{I}}{\partial A_*}\ , & i \int d^3x \frac{\partial \tilde{f}}{\partial \eta_*} &= \frac{\partial \tilde{I}}{\partial \eta_*}    \ .
\end{align}
We denote the solutions of these equations by $\boldsymbol{\Gamma}_{*ij}$, $\boldsymbol{A}_*$ and $\boldsymbol{\eta_*}$. They are functions of the original data, $h_{ij}$ and $\chi$. The semi-classical approximation thus gives the following relation between the two extremizing actions,
\begin{align}
I_{\textrm{extr}}[h_{ij},\chi] &= \tilde{I}_{\textrm{extr}}[\boldsymbol{\Gamma}_{*ij}, \boldsymbol{A}_*, \boldsymbol{\eta_*}] - i\int{d^3x\ \tilde{f}(\boldsymbol{\Gamma}_{*ij}, \boldsymbol{A}_*, h_{ij}, \chi, \boldsymbol{\eta_*})}  \ .   \label{eqn:transformedAction}
\end{align}

One can use this to determine the behavior of the on-shell actions in the large volume limit. The asymptotic behavior of the action $I_{\textrm{extr}}(h_{ij}, \chi)$ in terms of the asymptotic expansions given in \eqref{eqn:asymptoticAnsatz} is known to be
\begin{align}
I_{\textrm{extr}} &\approx \frac{i}{\kappa} \int{ \D^3x\ \sqrt{\gamma}\left( \frac{2}{\eta_*^3}-\frac{R_{(\gamma)}}{4\eta_*} - \frac{\kappa}{4\sigma} \left[ \alpha^2 \sigma^2 \gamma^{\lambda_-/\sigma} \eta_*^{-\sigma} - \beta^2 \gamma^{-\lambda_+/\sigma} \eta_*^{\sigma} \right]+\mathcal{O}(\eta_*) \right) } + I_\text{IR} \ ,
\end{align}
where $I_\text{IR}$ is an $\eta_*$-independent constant that depends on the non-asymptotic behavior of the on-shell action, i.e. on the value of the fields around the SP. Hence the diverging terms in the on-shell action $I$ are equal to those of the generating function \eqref{eq:generatingFunction}. This means there are no diverging terms left in $\tilde I$: the on-shell action is regulated by the canonical transformation to asymptotically meaningful coordinates, as expected\footnote{This can also be seen by writing $\tilde{I}_{\text{extr}} = -i \int{(\Pi_{(\Gamma)}^{ij} \D \Gamma_{ij} + B \D A + \sqrt{\Gamma} \tilde{H} \D \eta / \eta)}$. The Hamiltonian $\tilde{H}$ vanishes on-shell, as a result of the Hamiltonian constraint, and the other terms in the on-shell action remain finite, by virtue of the finiteness of the asymptotic variables. Therefore, $\tilde{I}_{\text{extr}}$ cannot diverge.}.
This is of course consistent with the usual counterterms employed in dS/CFT, which are given by
\begin{align}
S_{ct}[h,\chi]&=-\frac{2i}{\kappa} \int d^3 x \sqrt{h} + \frac{i}{2\kappa} \int d^3 x \sqrt{h}\ ^{(3)}R + \frac{i\lambda_-}{2} \int d^3x \sqrt{h}\ \chi^2 +\ldots
\end{align}
where the dots refer to additional scalar counterterms that enter for certain scalar masses only. The divergent parts of the counterterms $S_{ct}$ are equal to those of $f$.

\subsection{Implications for dS/CFT}\label{sec:asympHolography}
		
The relation between the semi-classical actions in \eqref{eqn:transformedAction} is reminiscent of the relation \eqref{eqn:holographicAction} between the action computed in the dS and in the AdS representations of the NBWF saddle points and leads, in the large volume limit, to the following chain of equalities,
\begin{align}
\tilde{I}_{\text{extr}}[\Gamma_{*ij}, A_*, \eta_*] = I_{\text{extr}}[h_{ij}, \chi] - i S_{ct} = -I_{aAdS}^{\text{reg}}[\bar{\gamma}_{*ij}, \bar{\alpha}_*]  \ ,
\end{align}
where $\bar{\gamma}_{ij}$ and $\bar{\alpha}$ are the natural variables from an AdS viewpoint, defined below \eqref{eqn:holographicAction}, and thus {\it locally related} to the argument of the wave function. Using \eqref{eq:dSCFT} this leads to the following formulation of a semiclassical dS/CFT correspondence 
\begin{align}
\tilde{\Psi}_{as}(\gamma_{ij}, \alpha) = \frac{1}{Z_{QFT}(\bar{\gamma}_{ij}, \bar{\alpha})}  \ , \label{eq:conjecture}
\end{align}
where we remind the reader that $Z_{QFT}$ is the partition function of a deformation of a Euclidean AdS/CFT dual.
This is both a more elegant and a cleaner formulation of dS/CFT than \eqref{eq:dSCFT}, since it is stated purely in terms of quantities that are available in the dual QFT. There is no need to involve the local counterterms in the bulk. Evaluating the wave function at finite scale factor rather than asymptotically loosely corresponds to adding a UV regulator $\upsilon$ in the boundary theory.
In what follows we fix the overall scale of the boundary metric to have volume one. This is consistent with Vol($\Gamma_{ij}) \to 1$, as implied by our choice of $\eta_0$ above.

The variables with a bar differ from the original variables by a phase only. More specifically, $\bar{\gamma}_{ij}$ and $\bar{\alpha}$ are the coefficients that are real in the FG expansion in the asymptotic AdS domain. Whereas $\eta$ is real on the dS part of the contour, the radial AdS variable $z = -i \eta$. Using the analyticity of \eqref{eqn:asymptoticAnsatz}, one can write the FG expansions in terms of $z$ and the barred variables, with 
\begin{align} \label{complex}
\bar{\Gamma}_{ij} &= -\Gamma_{ij} \ , &   \bar{\Pi}_{(\Gamma)}^{ij} &= \Pi_{(\Gamma)}^{ij}    \ , &   \bar{A} &= e^{-i\pi \lambda_- \lambda_+ / \sigma} A \ , &   \bar{B} = e^{i\pi \lambda_- \lambda_+ / \sigma}B \ .
\end{align}
The expectation value of the CFT stress tensor is dual to the subleading fall-off of the metric and the bulk scalar field corresponds to the expectation value of the dual scalar operator,
\begin{subequations} \label{eq:dsDict} 
\begin{align}
\braket{\mathcal{T}^{ij}(\vec{x})} = \frac{\delta}{\delta \bar{\gamma}_{ij}} \log Z_{\textrm{QFT}} \approx \frac{\delta \tilde{I}_{\textrm{extr}}[\Gamma_{*ij}, A_*, \eta_*]}{\delta \Gamma_{*ij}} &\to \bar{\pi}_{(\gamma)}^{ij}(\vec{x})  \ ,  \\ 
\braket{\mathcal{O}(\vec{x})} = \frac{\delta}{\delta \bar{\alpha}} \log Z_{\textrm{QFT}} \approx -e^{i\pi \lambda_+ \lambda_- / \sigma} \frac{\delta \tilde{I}_{\textrm{extr}}[\Gamma_{*ij}, A_*, \eta_*]}{\delta A_*} &\to \bar{\beta}(\vec{x})  \ .
\end{align} 
\end{subequations}
Here $\bar{\pi}_{(\gamma)}^{ij}$ and $\bar{\beta}$ are the asymptotic values of $\bar{\Pi}_{(\Gamma)}^{ij}$ and $\bar{B}$ on the extremizing solution for $\upsilon \to i \infty$, i.e. $\eta_* \to 0$. 

The phase factors in the relations \eqref{complex} between the asymptotic coefficients on the dS and the AdS branches means the vevs in the dual are in general complex when the argument of the dS wave function is real. This sharpens the question whether the emergence of classical space-time evolution in the large volume limit along the dS branch can be understood from the dual partition function.

\section{Classicality 2.0}\label{sec:classicality}

The NBWF in its usual form in terms of the variables $(h_{ij},\chi)$ oscillates very rapidly in the large volume limit. The large phase factor means the classicality conditions \eqref{eq:classicalityIntro} hold almost automatically, so that the wave function predicts that $(h_{ij},\chi)$ evolves classically. In fact the classical behavior can be understood as a consequence of the Wheeler-DeWitt equation in the large volume regime and therefore applies to any wave function in terms of these variables that satisfies the Hamiltonian constraint.

By contrast the wave function $\tilde \Psi$ need not oscillate and has no exponent that diverges in the large volume limit. This leads to the question whether it obeys the classicality conditions \eqref{eq:classicalityIntro} at large volume. The derivation in \cite{Hartle2008} of the conditions required for a wave function to predict classical evolution is general and applies also to the wave function in terms of the new asymptotic variables. Asymptotically the classicality conditions in the latter formulation are
\begin{subequations} \label{eq:asymptoticClassicality}
\begin{align}
| \nabla_{A_*} \tilde{I}_R | = | \Im B_* (\vec x)| &\ll | \Re B_*(\vec x) | = |\nabla_{A_*} \tilde{S}| \ , \\
| \nabla_{\Gamma_{*ij}} \tilde{I}_R | = | \Im \Pi_{(\Gamma)}^{ij}(\vec x) | &\ll | \Re \Pi_{(\Gamma)}^{ij}(\vec x) | = | \nabla_{\Gamma_{*ij}} \tilde{S} |    \ .   
\end{align}
\end{subequations}
These conditions are stronger than the original classicality conditions \eqref{eq:classicalityIntro}, derived from the NBWF in its usual `bulk' form. In particular, the original conditions do not involve the subleading coefficients in the Fefferman-Graham expansion. From the dual viewpoint they involve the sources only whereas \eqref{eq:asymptoticClassicality} are requirements on the vevs. 

This means that the ensembles of classical histories predicted by both wave functions may not be identical. We investigate and confirm this in the next section in a minisuperspace approximation in which we can verify the classicality conditions and identify the ensemble explicitly. 

This difference does not point towards an inconsistency; certain variables may exhibit classical behavior when others don't. One may ask, however, given two different notions of classicality, which one is more physical? While the original classicality conditions are natural from the point of view of an observer in the bulk it is clear that holography suggests the new set of stronger conditions \eqref{eq:asymptoticClassicality} is in fact more accurate and appropriate in quantum gravity.

\section{Minisuperspace model} \label{sec:minisuperspace}

In this section we compute the ensemble of classical histories predicted by $\tilde \Psi$ in a minisuperspace model consisting of homogeneous and isotropic histories. Adhering to the notation introduced above, the saddle point geometries can be written as
\begin{align}
ds^2= d\tau^2 + a^2(\tau) d\Omega_3^2	\ ,
\end{align}
where $d\Omega_3^2$ is the metric on the unit three-sphere. As before we consider gravity coupled to a single scalar field described by the action \eqref{eq:I_E} in which we take the potential $V$ to be quadratic. The usual NBWF is therefore a function of the boundary values $(b,\chi)$ of the scale factor $a(\tau)$ and scalar field $\phi(\tau)$. 

In \cite{Hartle2008} the usual semiclassical NBWF in this minisuperspace approximation was evaluated by systematically solving for the saddle points with the boundary conditions that $a=b$ and $\phi=\chi$ at the boundary $\tau = \upsilon$ and that geometry and field are regular at the SP of the instanton where the scale factor vanishes, say at $\tau=0$. 

The boundary conditions mean the saddle point solutions are generically complex. As discussed in Section \ref{hol} their geometric representation depends on the choice of contour in the complex $\tau$-plane connecting the SP with the endpoint $\tau=\upsilon$. 
Along a commonly used `dS contour' the analysis of \cite{Hartle2008} identified a one-parameter set of saddle point geometries consisting of a slightly deformed Euclidean four-sphere that makes a smooth transition through a complex intermediate region, to a Lorentzian inflationary history in which the scalar field slowly rolls down to the minimum of its potential. Those saddle points were found to obey the usual classicality conditions in the large volume limit, leading \cite{Hartle2008} to conclude that the usual NBWF in this minisuperspace model describes a one-parameter family of inflationary universes that are asymptotically de Sitter. We now generalize this analysis and compare with the classical predictions of $\tilde \Psi$ at large volume.

\subsection{Minisuperspace wave function \texorpdfstring{$\tilde \Psi$}{tilde{Psi}}}

We first construct the minisuperspace wave function $\tilde \Psi$. The Fefferman-Graham expansions \eqref{eqn:asymptoticAnsatz} in the minisuperspace approximation reduce to 
\begin{align}
a &= \frac{\gamma}{\eta} + \frac{\eta}{4\gamma} - \frac{\kappa \alpha^2}{16 \pi^2} \gamma^{\frac9\sigma - 2} \eta^{2-\sigma} + \frac{\kappa \gamma_{(3)}}{36 \pi^2\gamma} \eta^2 -\frac{\kappa \beta^2}{16\pi^2\sigma^2} \gamma^{-\frac9\sigma - 2} \eta^{2+\sigma} + \order{\eta^3}    \ , \nonumber \\
\phi &= \frac{\alpha}{\sqrt{2}\pi} \gamma^{3\lambda_- / \sigma} \eta^{\lambda_-} - \frac{\beta}{\sqrt{2}\pi \sigma} \gamma^{-3\lambda_+ / \sigma} \eta^{\lambda_+} + \order{\eta^{\lambda_- + 1}}  \label{eqn:FGMSS}\ ,
\end{align}
where we have defined\footnote{We use in this section the same notation as in the previous sections for simplicity, but be aware that their meaning is not completely the same. For example $\gamma$ is not the determinant of $\gamma_{ij}$ anymore. The same goes for the other variables.} $\gamma_{ij} \equiv \gamma^2 \Omega_{ij}$. 
The momenta conjugate to $a$ and $\phi$ can be found from $\pi_{(a)} = -12\pi^2 i a a'/\kappa$ and $\pi_{(\phi)} = 2 \pi^2 i a^3 \phi'$. All coefficients in the expansions \eqref{eqn:FGMSS} are given in terms of $(\alpha, \beta, \gamma, \gamma_{(3)})$ by the equations of motion. As before we define time dependent functions $\Gamma(\eta)$, $\Pi_{(\Gamma)}(\eta)$, $A(\eta)$ and $B(\eta)$ such that\footnote{$\gamma$ can be fixed to one by an appropriate choice of $\eta_0$ in \eqref{eqn:eta}. Furthermore, $\pi_{(\gamma)}$ is related to $\gamma_{(3)}$ as before.} $\Gamma \to 1$, $\Pi_{(\Gamma)} \to \pi_{(\gamma)}$, $A \to \alpha$ and $B \to \beta$ for $\eta \to 0$. This allows us to write the expansions as a finite polynomial. The symplectic form \eqref{eqn:cantransf} becomes
\begin{align}
\pi_{(a)} \D a + \pi_{(\phi)} \D\phi + H \frac{\D\eta}\eta &= \Pi_{(\Gamma)} \D\Gamma + B \D A + \tilde{H} \frac{\D\eta}\eta + \D f + \mathcal{P}(\eta)    \ , \label{eqn:fullcantransf}
\end{align}
where $\mathcal{P}$ contains higher order terms in $\eta$ and where the new Hamiltonian $\tilde{H}$ and generating function $f$ are given by 
\begin{align}
\tilde{H} &= H - \Gamma\ \Pi_{(\Gamma)} + \frac{2m^2}\sigma A B  \ , \\
f &= -\frac{4\pi^2\Gamma^3}{\kappa\eta^3} + \frac{3\pi^2\Gamma}{\kappa\eta} + \frac{\lambda_-}{\sigma} AB - \frac{\Gamma\Pi_{(\Gamma)}}3 + \frac\sigma4 \Gamma^{9/\sigma} A^2 \eta^{-\sigma} - \frac1{4\sigma} \Gamma^{-9/\sigma} B^2 \eta^{\sigma}   \ ,
\end{align}
and the wave function $\tilde \Psi$ in terms of the asymptotic variables can be written as
\begin{align}
\tilde{\Psi}(\Gamma_*,A_*,\eta_*) &= \int_{\mathcal{C}}{\cD \Gamma \cD \Pi_{(\Gamma)} \cD A \cD B \ e^{i \int{B \D A} +i\int{\Pi_{(\Gamma)} \D \Gamma}+i \int{\frac{\D\eta}{\eta} \tilde{H}(\Gamma,\Pi_{(\Gamma)},A,B)}}} \ ,
\end{align}
where $\tilde{f}$ is obtained from $f$ by substituting the momenta $(B, \Pi_{(\Gamma)})$ in terms of the coordinates $(a, \phi, A, \Gamma)$, yielding
\begin{align}
\tilde f =& \frac{\pi^2 \Gamma_*^3}{\eta_*^3} \left( \frac8{\kappa} - \lambda_+\chi^2 \right) - \frac{12 \pi^2b \Gamma_*^2}{\kappa \eta_*^2} + \sigma A_* \chi \sqrt{2}\pi\Gamma_*^{3\lambda_+ / \sigma} \eta_*^{-\lambda_+} + \frac{6\pi^2\Gamma_*}{\kappa \eta_*} - \frac\sigma2 A_*^2 \Gamma_*^{9/\sigma} \eta_*^{-\sigma} \ ,
\end{align}
and where $\tilde{H}$ now includes the higher order terms in $\eta$ coming from $\mathcal{P}$, namely
\begin{align}
\mathcal{P}(\eta) =& - \eta^2 \left( \frac{\Pi_{(\Gamma)}}{12 \Gamma^2} + \frac3{16} (\sigma - 1) A^2 \Gamma^{9/\sigma - 3} \eta^{-\sigma} - \frac3{16 \sigma^2} (\sigma + 1) B^2 \Gamma^{-9/\sigma - 3} \eta^\sigma \right) \D \Gamma \nonumber \\
& - \frac{\kappa}{1728 \pi^2 \sigma^3} \Gamma^{-4-18/\sigma} \eta^{3-2\sigma} \left( -4 \Gamma \Pi_{(\Gamma)} - 9 (\sigma - 1) A^2 \Gamma^{9/\sigma} \eta^{-\sigma} + 9 (\sigma + 1) B^2 \Gamma^{-9/\sigma} \eta^{\sigma} \right) \nonumber \\
& \times \left( 4 \sigma \Gamma (\Gamma \D \Pi_{(\Gamma)} - \Pi_{(\Gamma)} \D \Gamma) + 9 A \Gamma^{9/\sigma} \eta^{-\sigma} ((2\sigma - 9) A \D \Gamma - 2 \sigma \Gamma \D A) \right. \nonumber \\
&\qquad \left. + \frac9{\sigma^2} B\Gamma^{-9/\sigma} \eta^{\sigma} ((2\sigma + 9) B \D \Gamma - 2 \sigma \Gamma \D B) \right)    \ .
\end{align} 

Finally the relation between the different on-shell actions in the semi-classical approximations becomes
\begin{align}
I_\text{extr}(b, \chi) = \tilde{I}_\text{extr}(\boldsymbol{\Gamma_*}, \boldsymbol{A_*}, \boldsymbol{\eta_*}) - i \tilde f( \boldsymbol{\Gamma_*}, \boldsymbol{A_*}, b, \chi, \boldsymbol{\eta_*}) \ .
\end{align}
The minisuperspace approximation allows for an explicit calculation of the actions which shows that all divergences of $I_\text{extr}$ are indeed contained in $\tilde f$.

\subsection{Classical Predictions}
Our main motivation to evaluate $\tilde \Psi$ in the minisuperspace approximation is to explicitly verify the differences between the classical predictions of both formulations of the NBWF implied by the classicality conditions \eqref{eq:classicalityIntro} expressed in terms of $(h_{ij},\chi)$ or \eqref{eq:asymptoticClassicality} terms of $(\gamma_{ij},\alpha)$.

Applied to the minisuperspace model, the classicality conditions in terms of the asymptotic variables \eqref{eq:asymptoticClassicality} require that asymptotically classical universes must have real $\beta$ and real $\gamma_{(3)}$. The latter condition follows from the former, because the Hamiltonian constraint implies $\Gamma \Pi_{(\Gamma)} = 2m^2 AB / \sigma$. Since $\Gamma_*$ and $A_*$ are real , $\Pi_{(\Gamma *)}$ has to be real if $B_*$ is real, and vice versa.

To evaluate $\tilde \Psi$ we first solve the equations of motion subject to the above (no-)boundary conditions. The saddle points can be viewed as solutions $\left( a(\tau),\phi(\tau) \right)$ in the complex $\tau$-plane, as discussed in Section \ref{hol}. Following \cite{Hartle2008} we label different solutions by the absolute value $\phi_0$ of the scalar field at the SP which we write as $\phi(\tau=0) = \phi_0 e^{i\theta}$. It follows from the FG expansions that in saddle point solutions associated with asymptotically classical histories, field and geometry become real along a vertical line somewhere in the $\tau$-plane, $\tau = x_\text{TP} + i y$. In \cite{Hartle2008} it was found that for each value of $\phi_0$, there is a single value of $\theta$ as well as a vertical line labeled by $x_\text{TP}$ along which the solutions become real and Lorentzian, satisfying the classicality conditions. 

\begin{figure}[ht]
\centering
\begin{subfigure}[b]{0.32\textwidth}
\includegraphics[width=\textwidth]{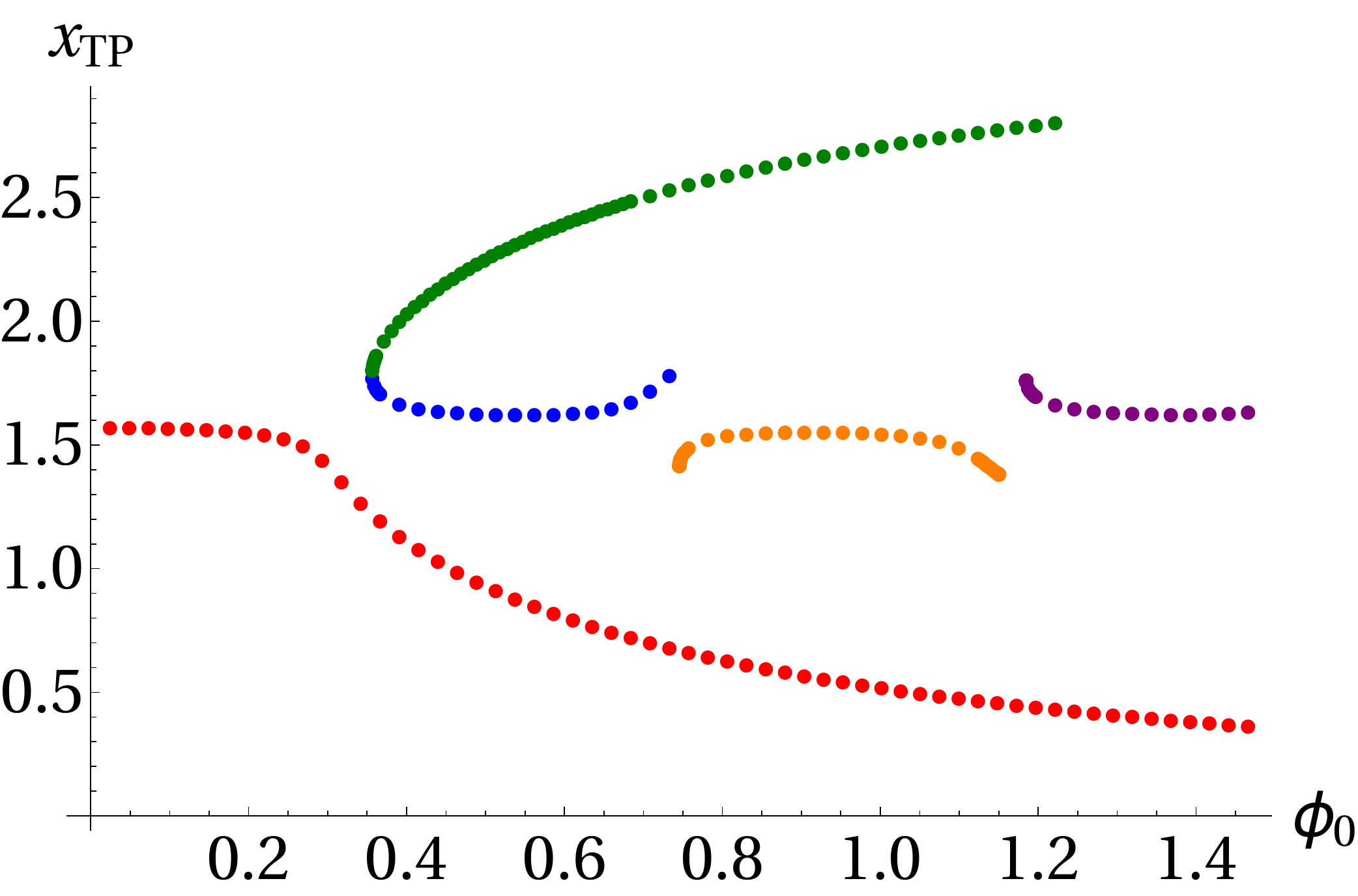}
\end{subfigure}
\begin{subfigure}[b]{0.32\textwidth}
\includegraphics[width=\textwidth]{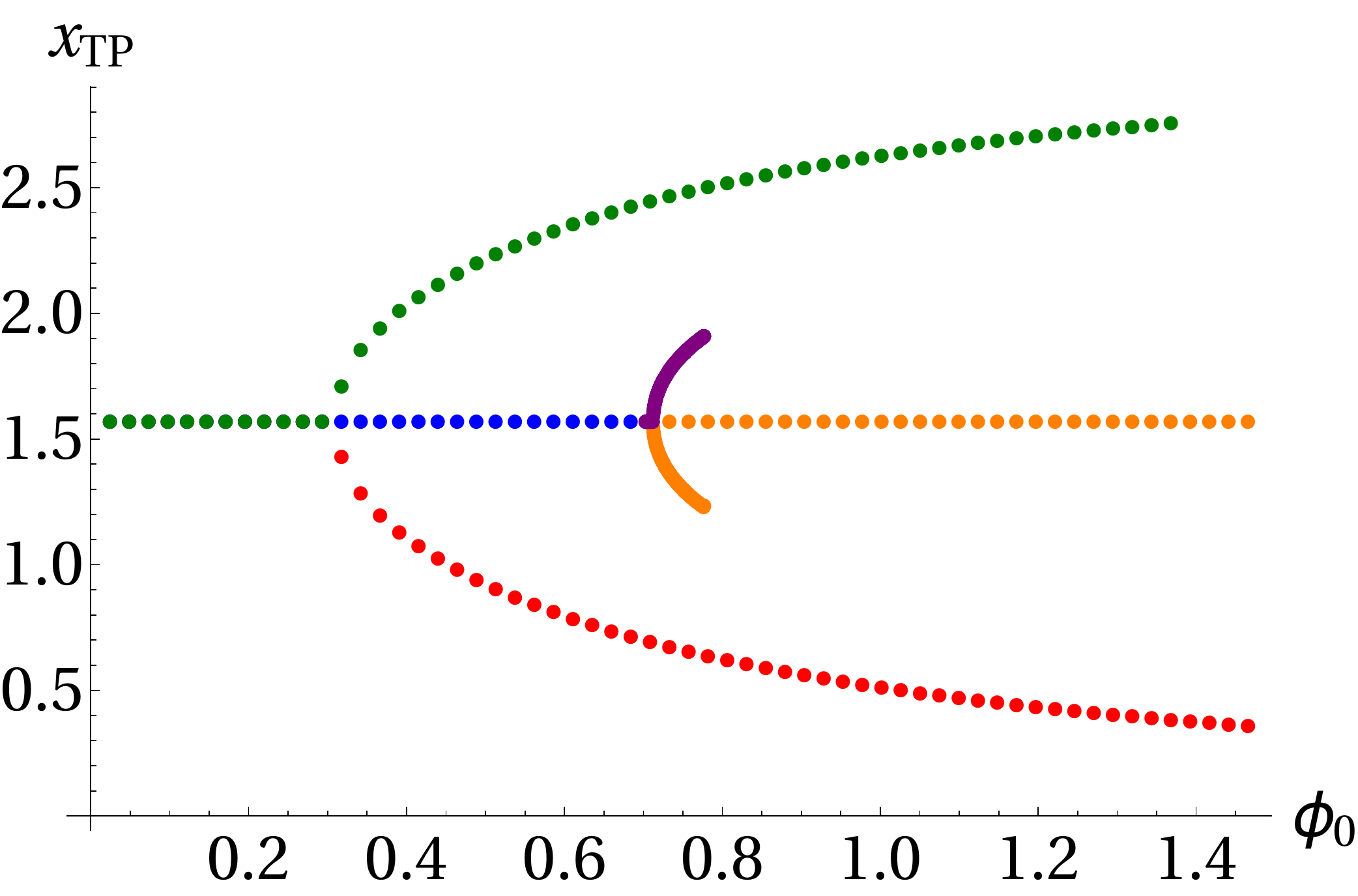}
\end{subfigure}
\begin{subfigure}[b]{0.32\textwidth}
\includegraphics[width=\textwidth]{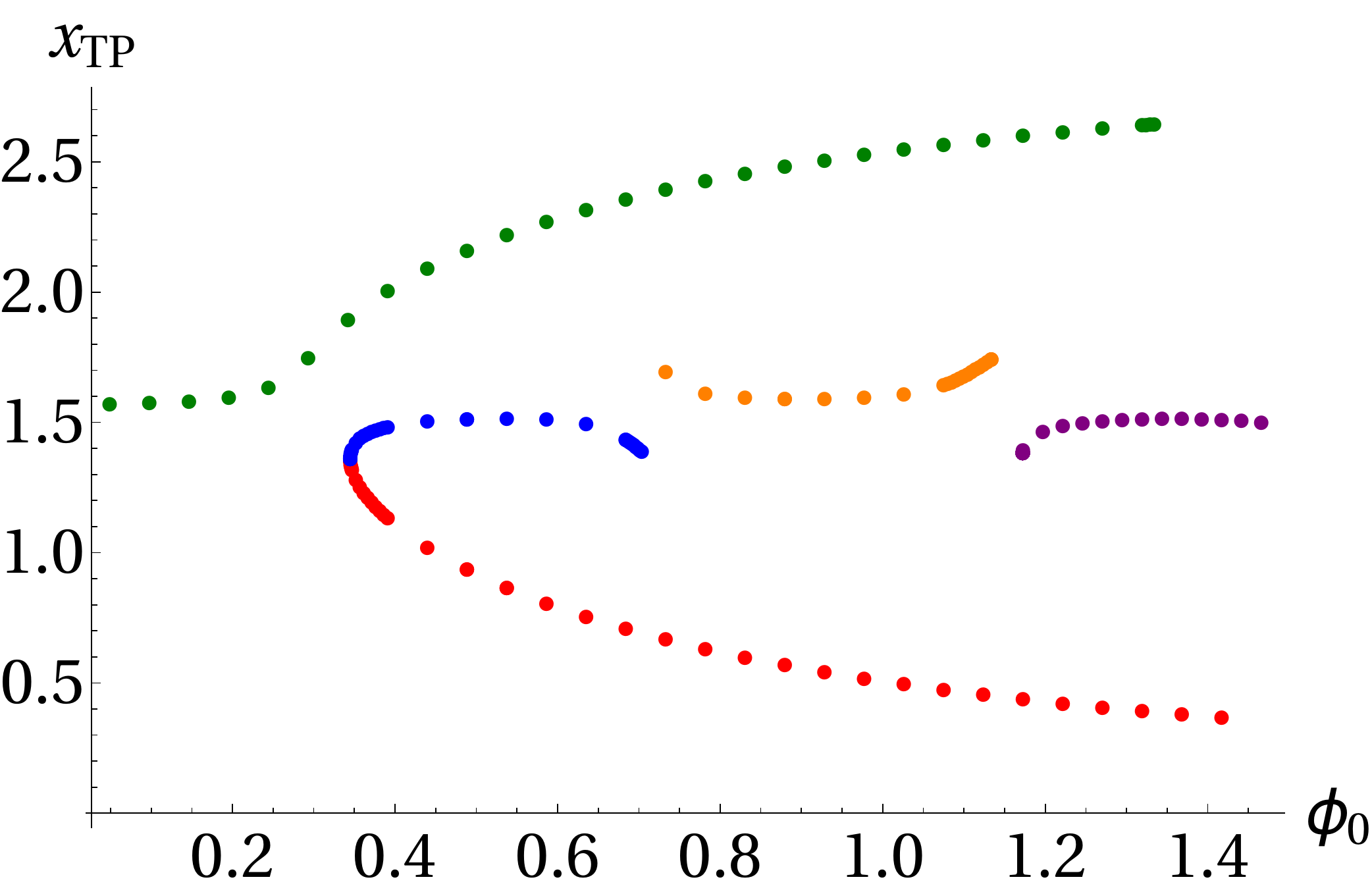}
\end{subfigure}\\
\begin{subfigure}[b]{0.32\textwidth}
\includegraphics[width=\textwidth]{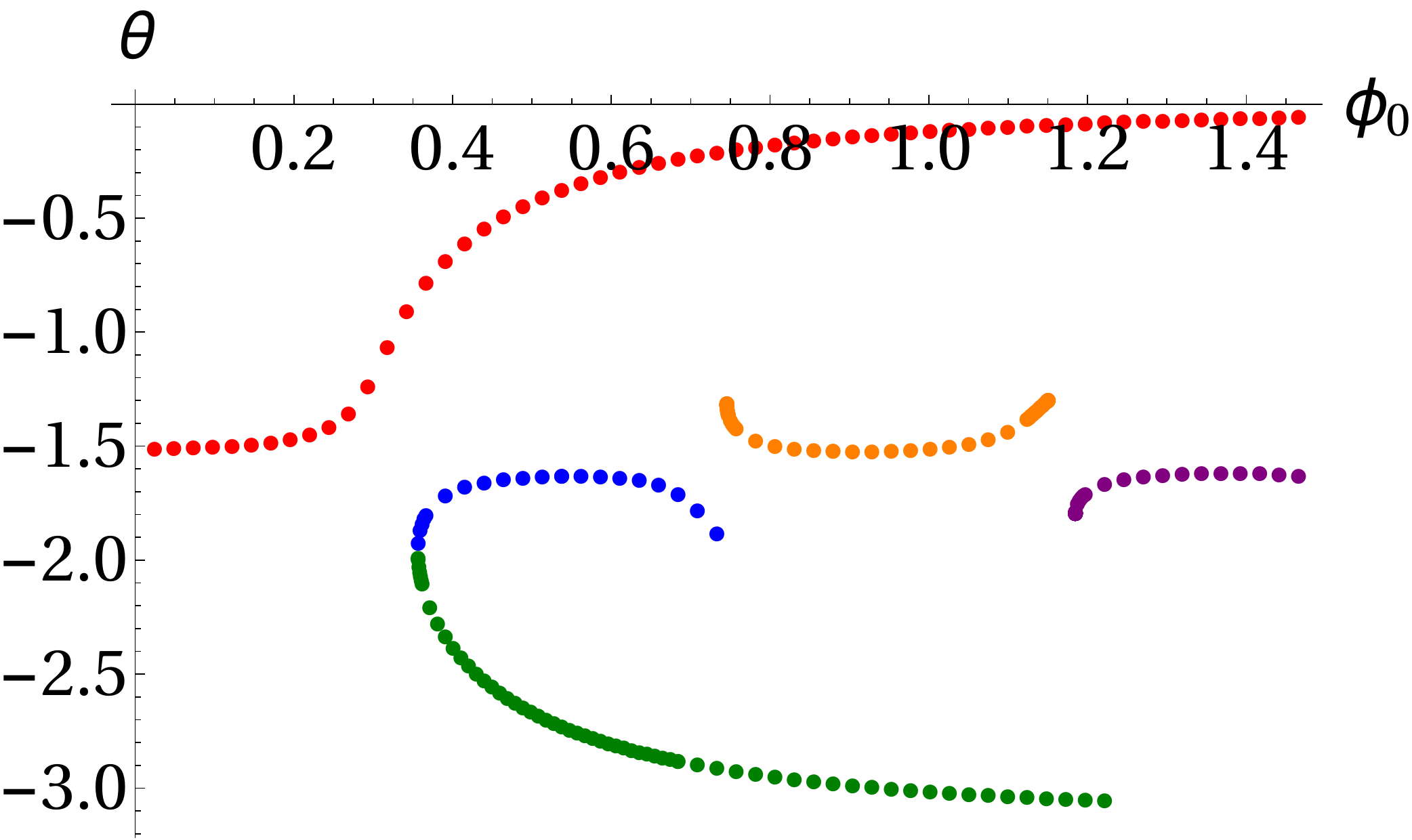}
\caption{$m=1.4$} \label{fig:bifurcationmu14}
\end{subfigure}
\begin{subfigure}[b]{0.32\textwidth}
\includegraphics[width=\textwidth]{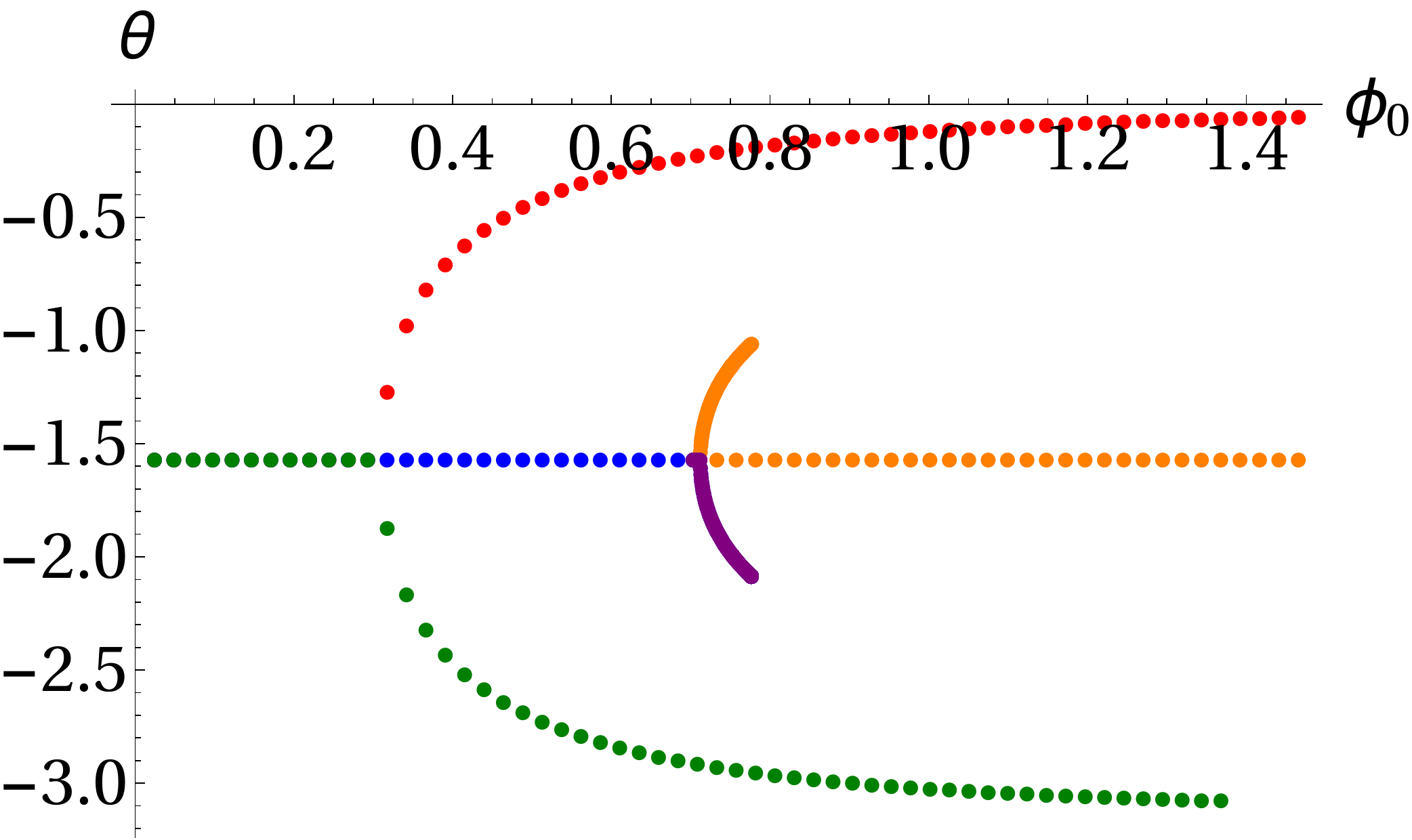}
\caption{$m=\sqrt{2}$} \label{fig:bifurcationmusqrt2}
\end{subfigure}
\begin{subfigure}[b]{0.32\textwidth}
\includegraphics[width=\textwidth]{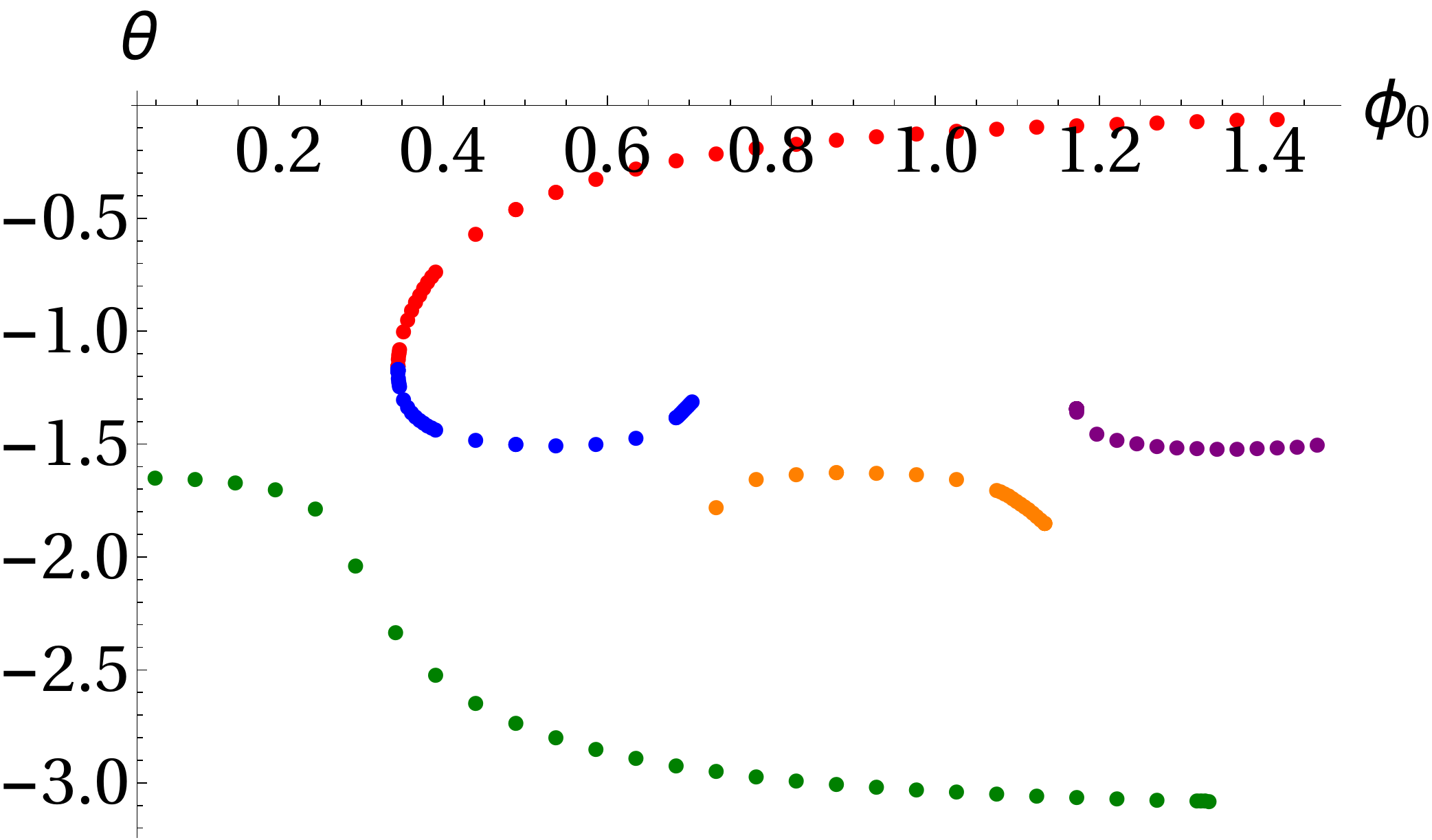}
\caption{$m=1.43$} \label{fig:bifurcationmu143}
\end{subfigure}
\caption{The semi-classical contributions to the NBWF in the minisuperspace approximation can be characterized by three numbers $\phi_0$, $\theta$ and $x_\text{TP}$ defined in the text. This figure shows the values of $\theta$ and $x_\text{TP}$ as a function of $\phi_0$ for which both $a(\tau)$ and $\phi(\tau)$ become real and the solutions asymptote to de Sitter space as $y \rightarrow \infty$. The set of solutions depends on the scalar mass. We show the solutions for three values of $m$; from left to right $1.4$, $\sqrt{2}$ and $1.43$.} \label{fig:bifurcation}
\end{figure}

We have performed a more systematic analysis of the classical predictions which we summarize in Figure \ref{fig:bifurcation} where we show, for three values of the scalar mass $m^2$, the values $(x_\text{TP}, \theta)$ in function of $\phi_0$ for which the saddle points become real at large times $y$. As can be seen, there are in general multiple one-parameter families of solutions. This generalizes the results obtained in \cite{Hartle2008}, where only the solutions depicted in red were found\footnote{Our result is consistent with observations in dimensions other than four \cite{Liu:2014ths} and for different potentials \cite{Battarra:2014kga}.}.

The parameters specifying the solutions change continuously with the value of the scalar field mass. The conformally coupled scalar with $m^2 = 2$ is a special value for which the space of solutions has an enhanced symmetry. This can be explained analytically, as described in Appendix \ref{subs:bifurcate}.

\begin{figure}[ht]
\centering
\begin{subfigure}[b]{0.32\textwidth}
\includegraphics[width=\textwidth]{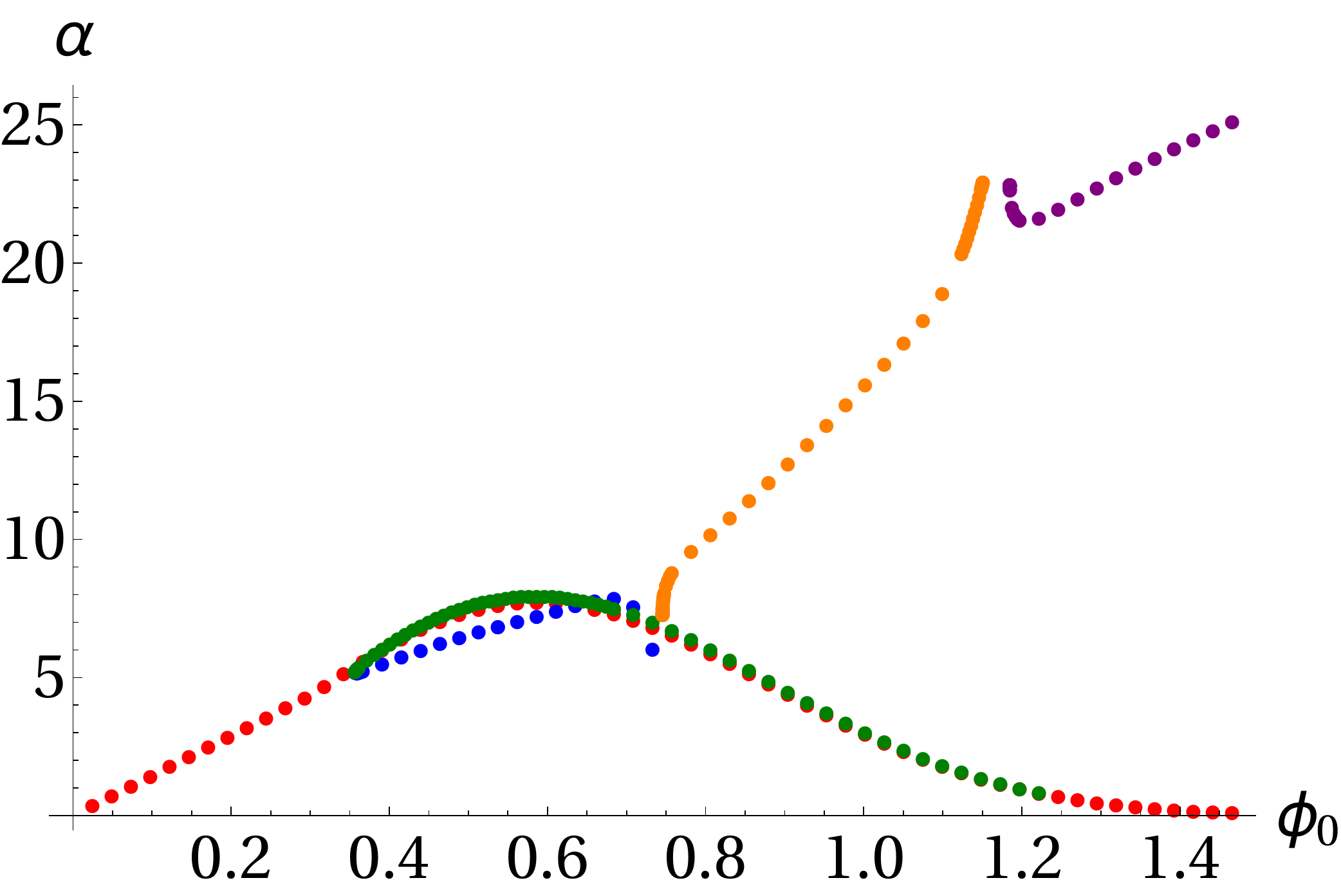}
\caption{$m=1.4$} \label{fig:alphavsphi0mu14}
\end{subfigure}
\begin{subfigure}[b]{0.32\textwidth}
\includegraphics[width=\textwidth]{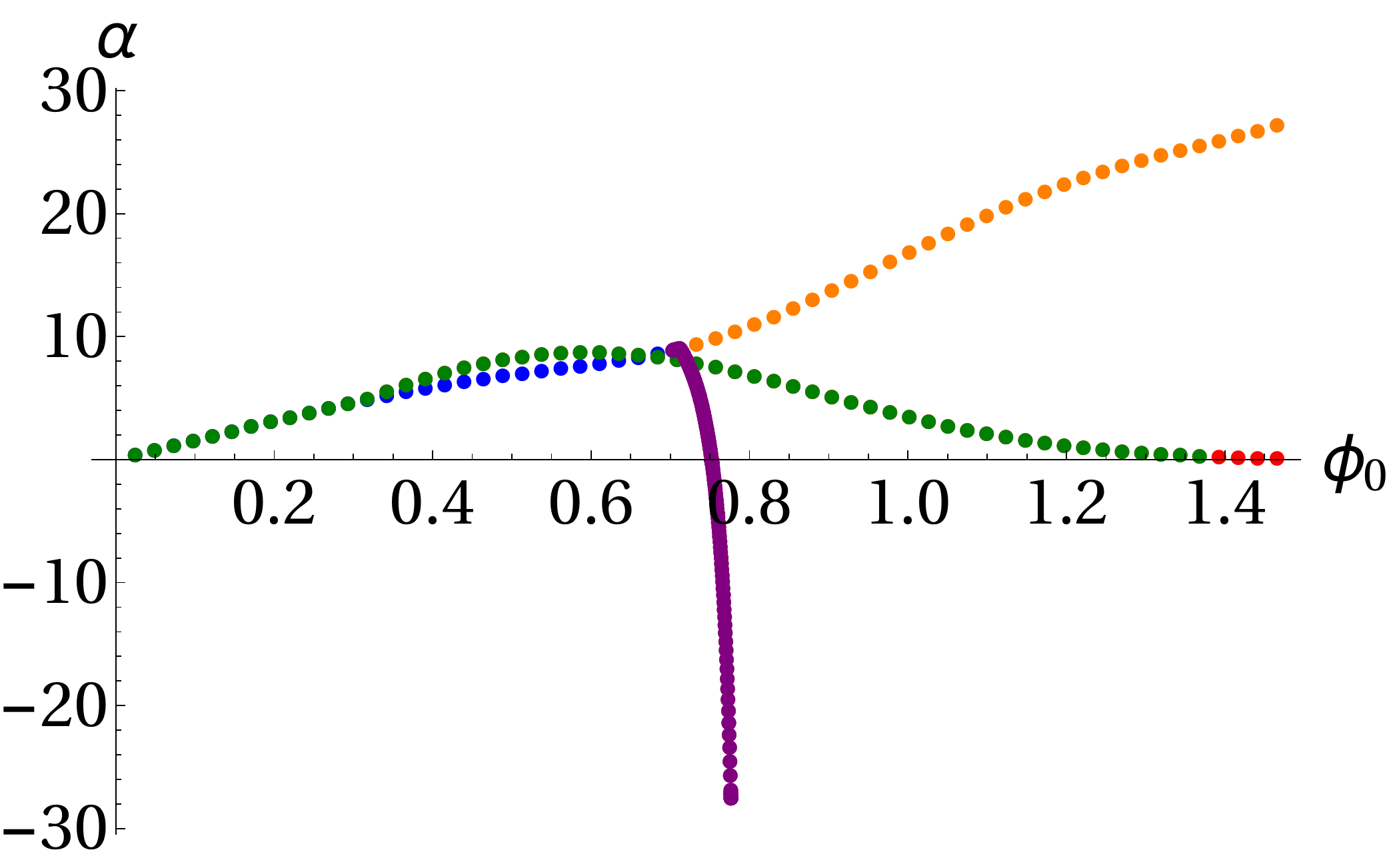}
\caption{$m=\sqrt{2}$} \label{fig:alphavsphi0musqrt2}
\end{subfigure}
\begin{subfigure}[b]{0.32\textwidth}
\includegraphics[width=\textwidth]{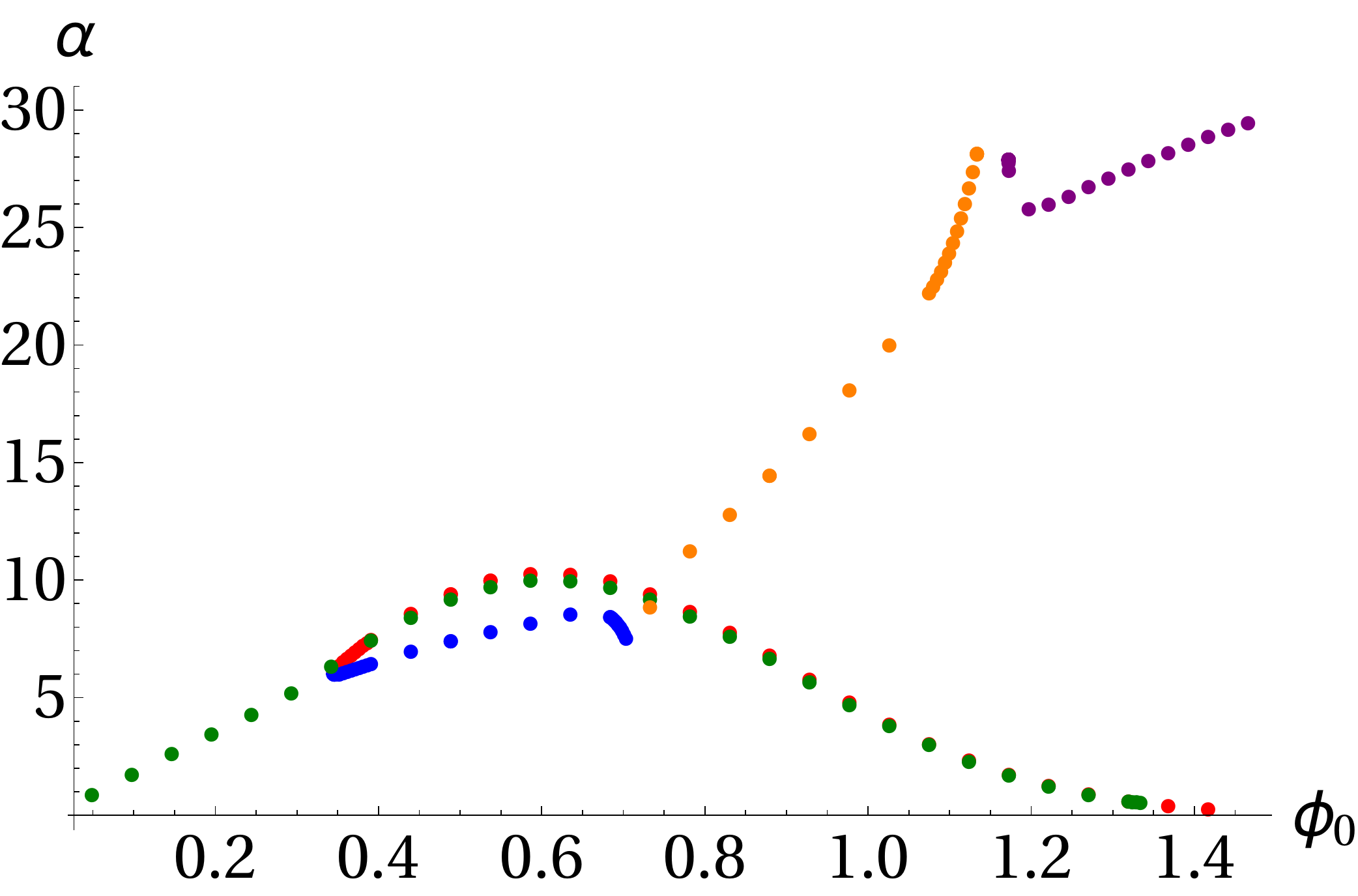}
\caption{$m=1.43$} \label{fig:alphavsphi0mu143}
\end{subfigure}
\caption{The value of $\alpha$ versus $\phi_0$ for the solutions shown in Figure \ref{fig:bifurcation}. The colors coincide with the colors used there. Notice that the red and green branches almost completely coincide} \label{fig:alphavsphi0} \end{figure}

With the solutions found above, the asymptotic wave function $\tilde{\Psi}$ can be constructed. It suffices to find the relation between the asymptotic parameters (dual to the sources) and the initial conditions denoted by $\phi_0$, in order to interpret the solutions above as saddle points of $\tilde \Psi$. In Figure \ref{fig:alphavsphi0} we show $\alpha$ as a function of $\phi_0$ for the solutions found in Figure \ref{fig:bifurcation}. This figure shows that the correspondence between $\phi_0$ and $\alpha$ is not one-to-one. Instead, it can happen that multiple saddle point solutions contribute to $\tilde{\psi}(\alpha)$ for a single value of $\alpha$, even within each of the "branches" identified in Figure \ref{fig:bifurcation}. We return to this point below.

\begin{figure}[ht]
\centering
\begin{subfigure}[b]{0.45\textwidth}
\includegraphics[width=\textwidth]{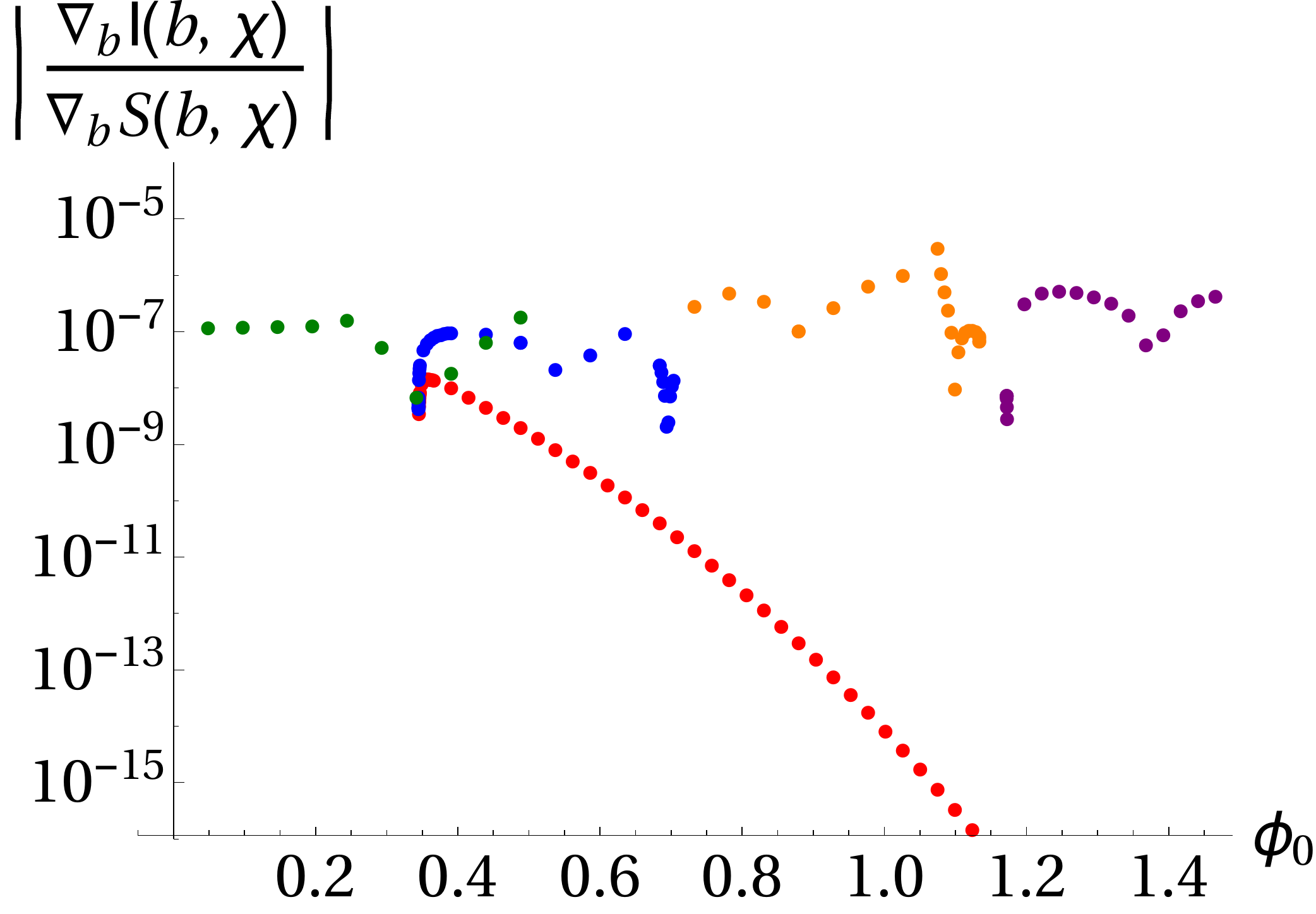}
\caption{In terms of the derivative w.r.t. $b$} \label{fig:classcondbmu143}
\end{subfigure}
\begin{subfigure}[b]{0.45\textwidth}
\includegraphics[width=\textwidth]{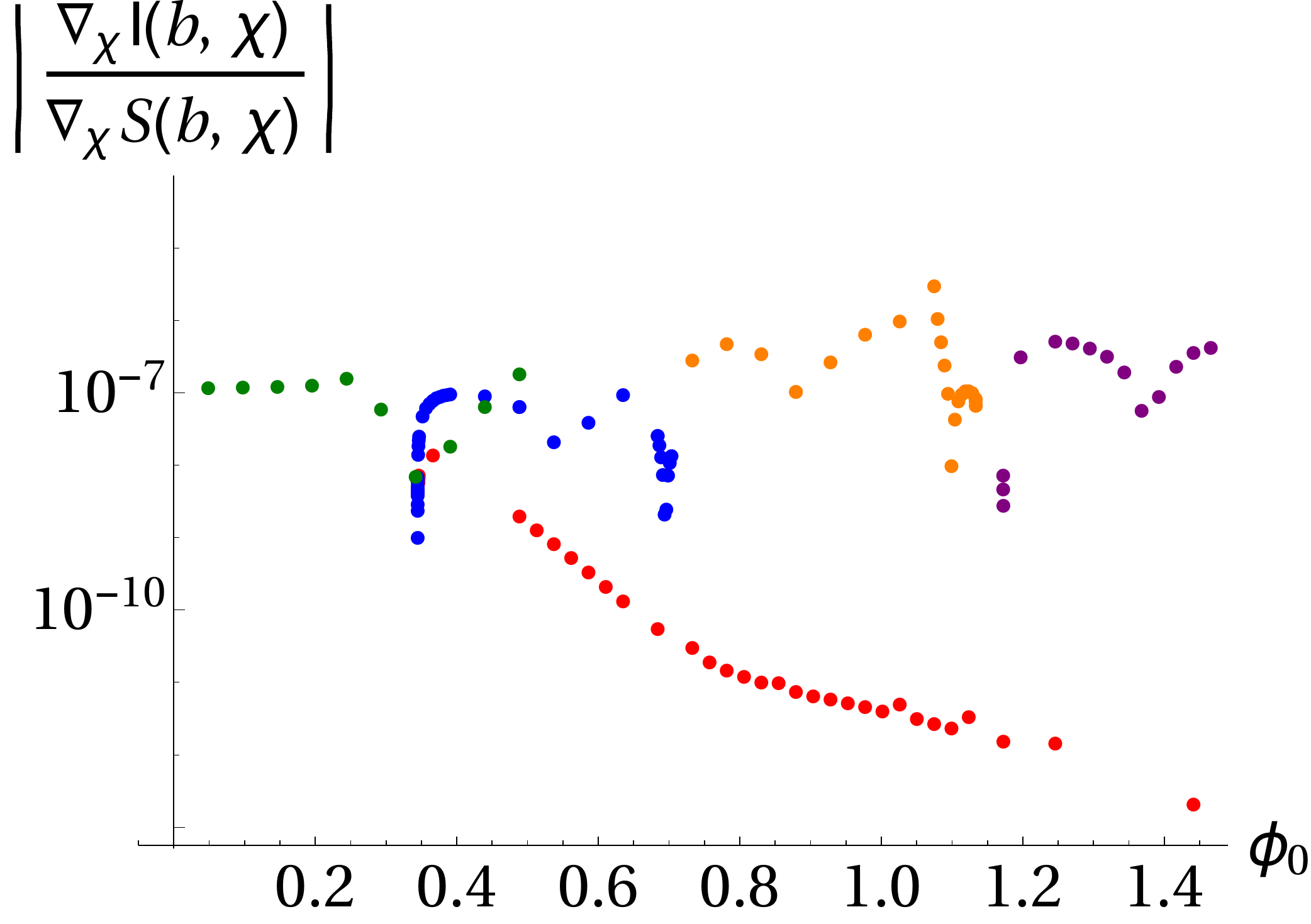}
\caption{In terms of the derivative w.r.t. $\chi$} \label{fig:classcondchimu143}
\end{subfigure}
\caption{The classicality conditions in terms of the variables $(b,\chi)$ for the scalar field solutions with $m=1.43$, evaluated at $\eta_* = e^{-6} \approx 2 \cdot 10^{-3}$. As expected, all solutions are predicted to behave classically asymptotically.} \label{fig:classcondchib}
\end{figure}

We now consider the classicality conditions, starting with their original formulation.
Figure \ref{fig:classcondchib} shows the ratio of the gradients of the real and imaginary parts of the action with respect to the usual variables $(b,\chi)$ for the solutions with $m=1.43$. The other values of the scalar mass give very similar results. If the ratio of the derivatives plotted in Figure \ref{fig:classcondchib} is small, the usual NBWF predicts that the corresponding homogeneous, isotropic configuration $(b,\chi)$ evolves classically at large volume. As anticipated in Section \ref{sec:classicality}, all solutions satisfy these classicality conditions. 

Notice that the values in these figures are of the expected order of magnitude. Because the imaginary part of the action goes as $S \sim \frac{1}{\eta_*^3}$ while $I$ remains of order 1, we expect that $| \nabla I / \nabla S | \sim \eta_*^3$, giving the resulting values of the classicality conditions. However, the semi-classical solutions indicated in red seem to do parametrically better than this. We will see that this distinction between the saddle points carries over to $\tilde \Psi$ and appears to be physically meaningful.

\begin{figure}[ht]
\centering
\begin{subfigure}[b]{0.32\textwidth}
\includegraphics[width=\textwidth]{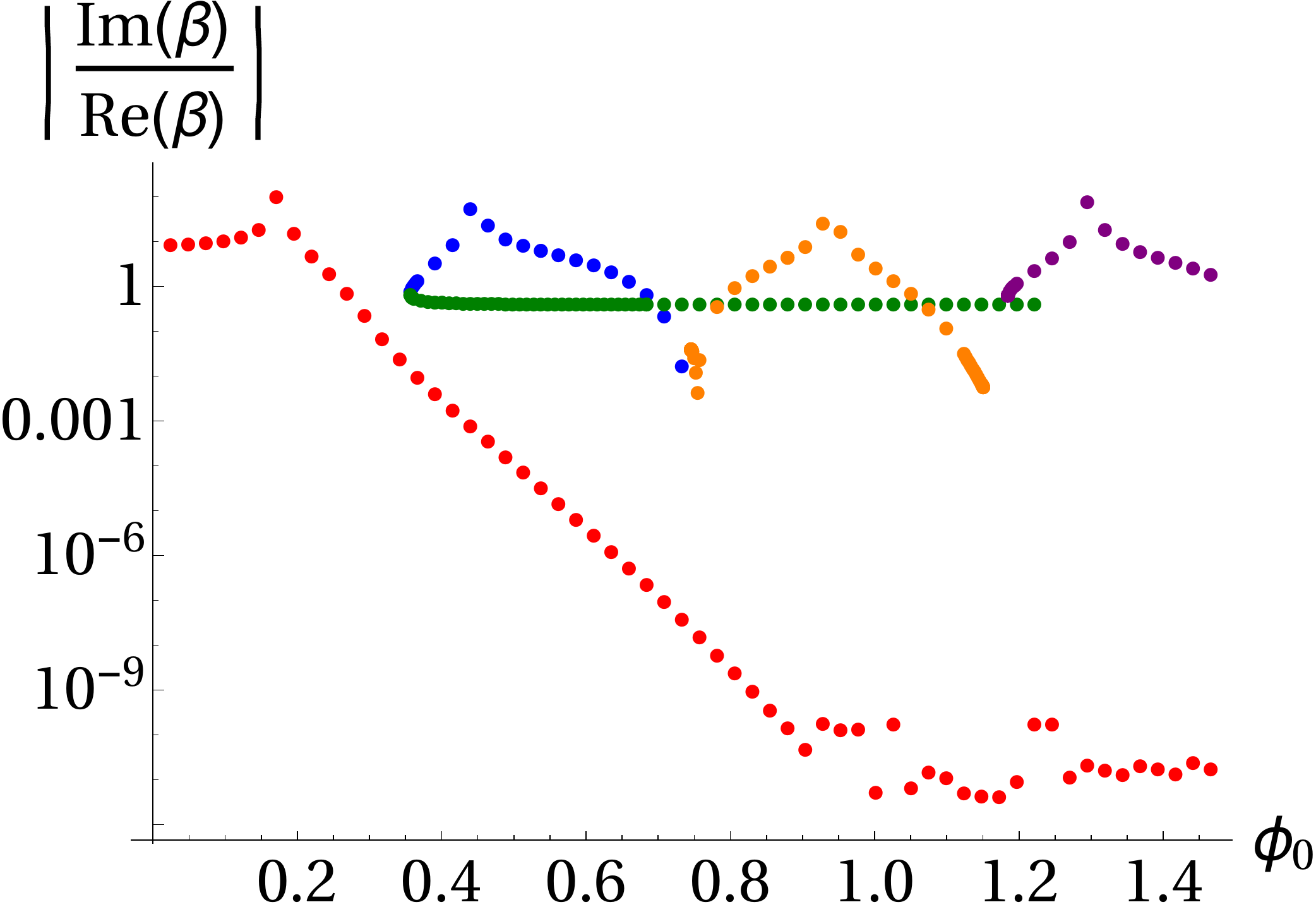}
\caption{$m=1.4$} 
\end{subfigure}
\begin{subfigure}[b]{0.32\textwidth}
\includegraphics[width=\textwidth]{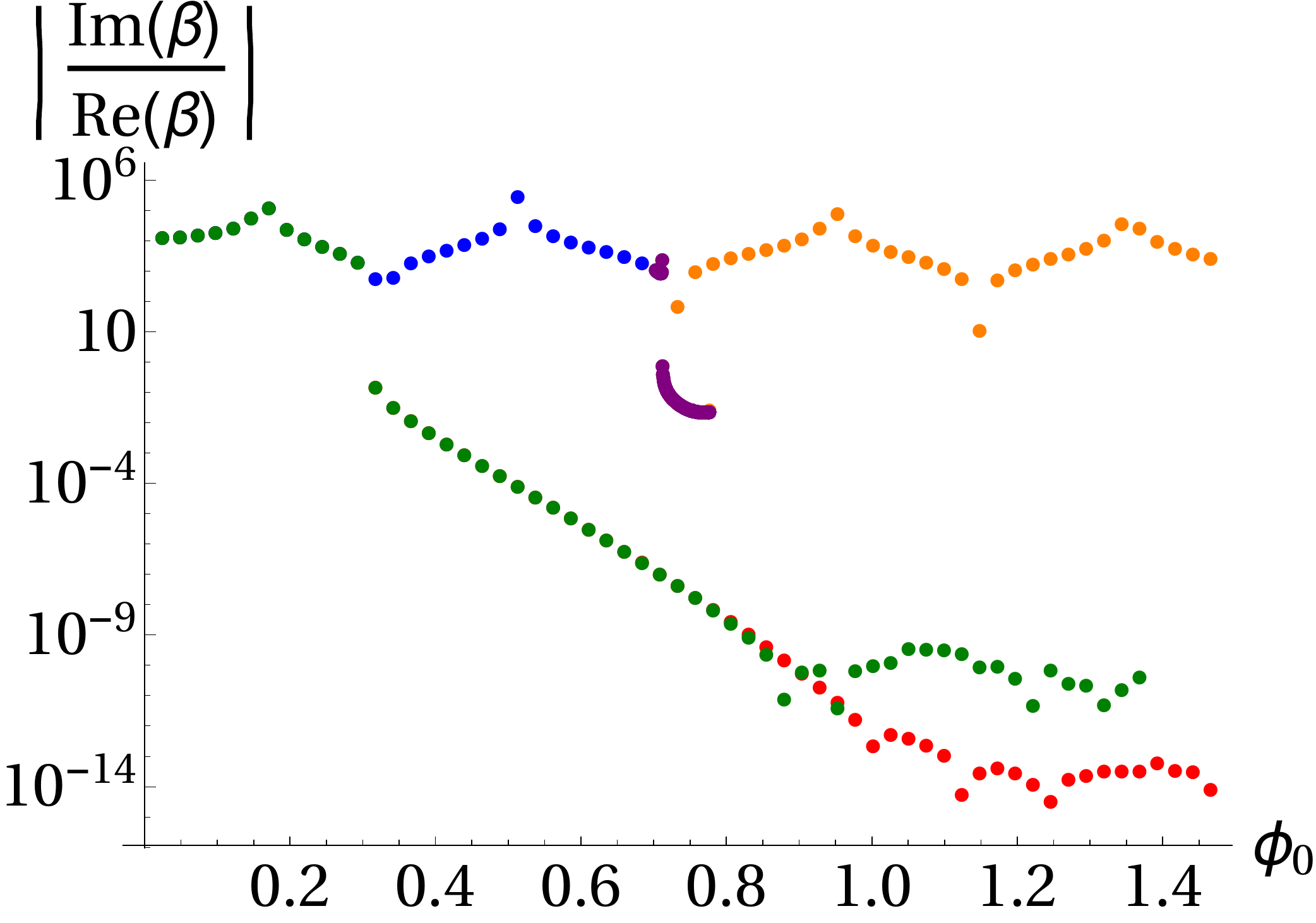}
\caption{$m=\sqrt{2}$} 
\end{subfigure}
\begin{subfigure}[b]{0.32\textwidth}
\includegraphics[width=\textwidth]{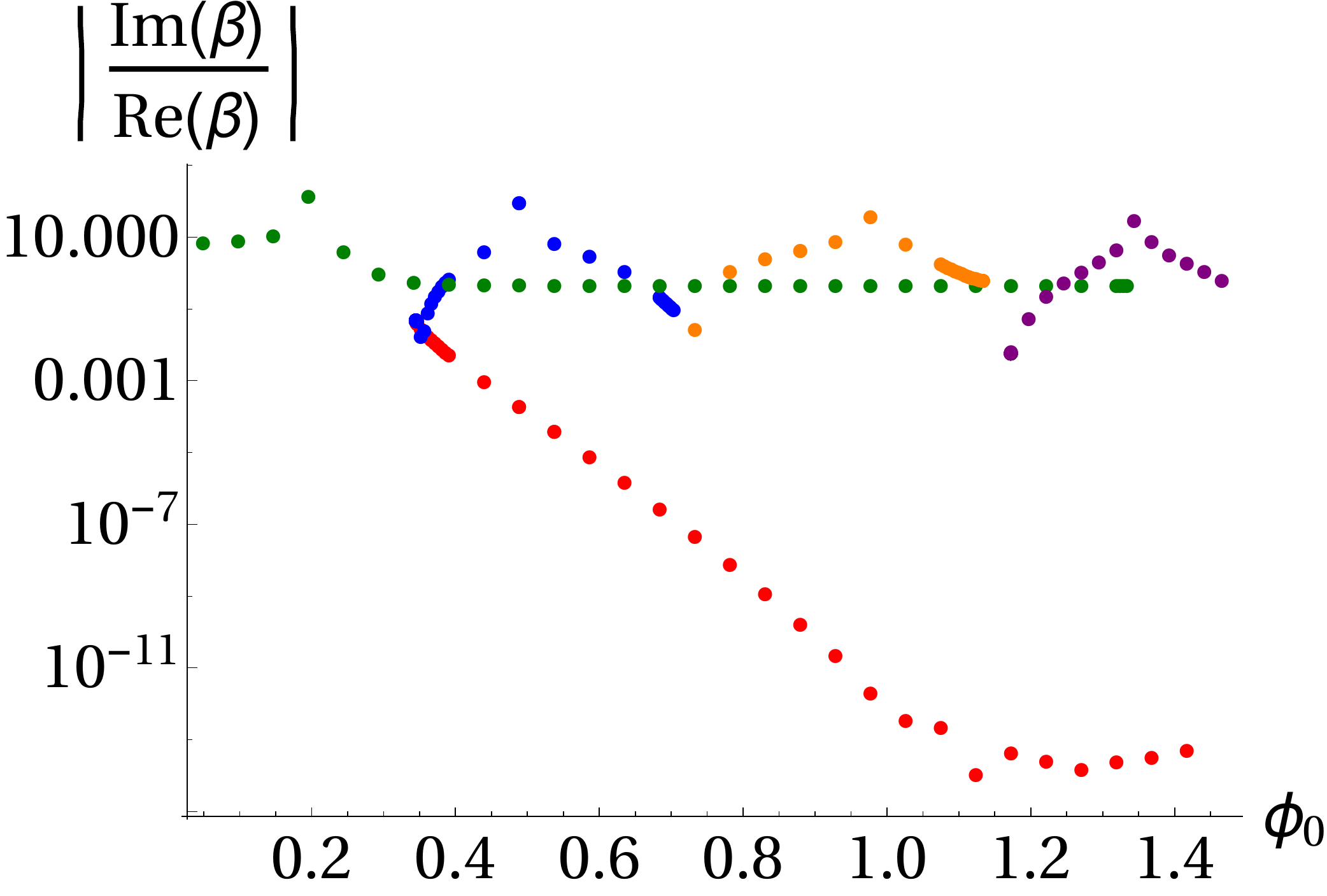}
\caption{$m=1.43$}
\end{subfigure}
\caption{The classicality conditions in terms of the asymptotic variables are shown. From left to right the ratios of the imaginary part of $\beta$ to its real part are plotted in function of $\phi_0$ for three different masses $1.4$, $\sqrt{2}$ and $1.43$. } \label{fig:classcondbeta}
\end{figure}

Next we evaluate the classicality conditions in terms of asymptotic variables. Figure \ref{fig:classcondbeta} shows the ratio of the imaginary part of $\beta$ to its real part as a function of $\phi_0$ for three different masses of the scalar field\footnote{From the holographic point of view, it would be more natural to plot these as functions of $\alpha$, but in this way it is easier to compare these results with the classicality conditions in bulk variables.}. A number of points are important. First, the scale on the $y$-axis is very different from the scale in Figure \ref{fig:classcondchib} -- the ratio is a lot larger. In fact, for some solutions the ratio is of order 1 or even larger. Hence these solutions do not obey the more stringent classicality conditions in terms of asymptotic variables. More precisely, none of the solutions that were previously unknown are classical. A remarkable observation is that for small $\phi_0$ there are no solutions at all that are predicted to behave classically, even though a perturbative analysis based on the original classicality conditions leads to the opposite conclusion \cite{Hartle2008}. Note also that for the conformal mass, the red and green branch coincide again and correspond to classical solutions. This was to be expected due to the symmetry of this case (see Appendix \ref{subs:bifurcate}). It is interesting to observe that a small breaking of this symmetry -- for example changing the mass from $\sqrt{2}$ to $1.4$ or $1.43$ -- has a drastic effect on the behavior of the green branch.

At a technical level, the discrepancy between the classical predictions of $\Psi$ and $\tilde \Psi$ can be traced to the generating function.
In particular, the comparison shows that the seemingly classical behavior of the new solutions in terms of the usual variables is entirely due to the growing phase factor in the usual NBWF, which is a universal surface term that is the generating function. This is absent in $\tilde \Psi$ -- and in the dual partition function -- and the new classicality conditions are not sensitive to this. Instead they are more stringent and depend on the interior dynamics and on the quantum state, which are encoded asymptotically e.g. in $\beta$.

For a history to behave truly classically it must obey both classicality conditions. It is not sufficient for classical behavior to be manifest in certain variables and not in others. Having said this, the prediction of classical behavior from a wave function of the universe remains inherently approximate. The classicality conditions \eqref{eq:asymptoticClassicality} are stated as an inequality allowing therefore for some quantum fuzz.

\begin{figure}[ht]
\centering
\begin{subfigure}[b]{0.32\textwidth}
\includegraphics[width=\textwidth]{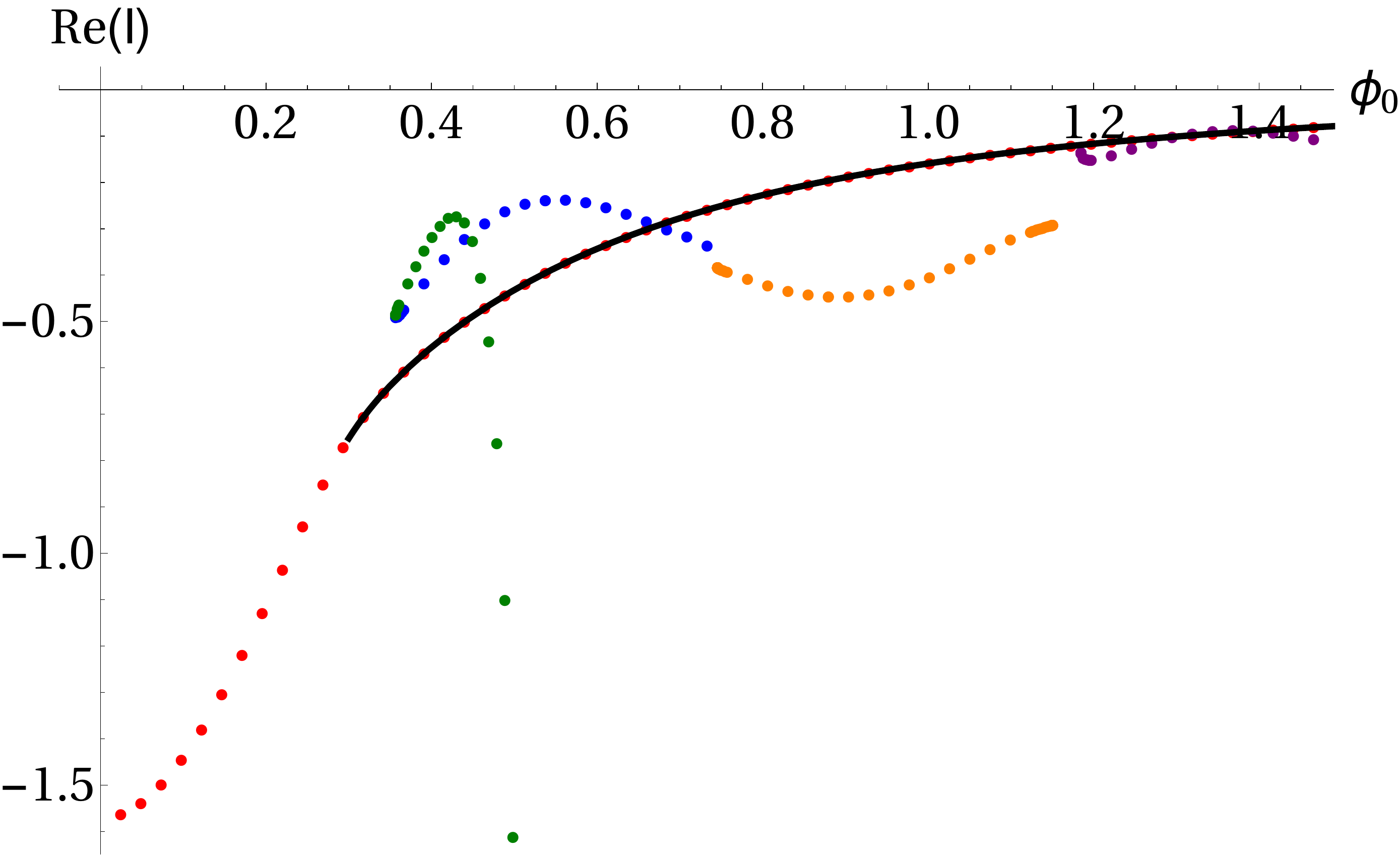}
\caption{$m=1.4$} \label{fig:bifurcationActionmu14}
\end{subfigure}
\begin{subfigure}[b]{0.32\textwidth}
\includegraphics[width=\textwidth]{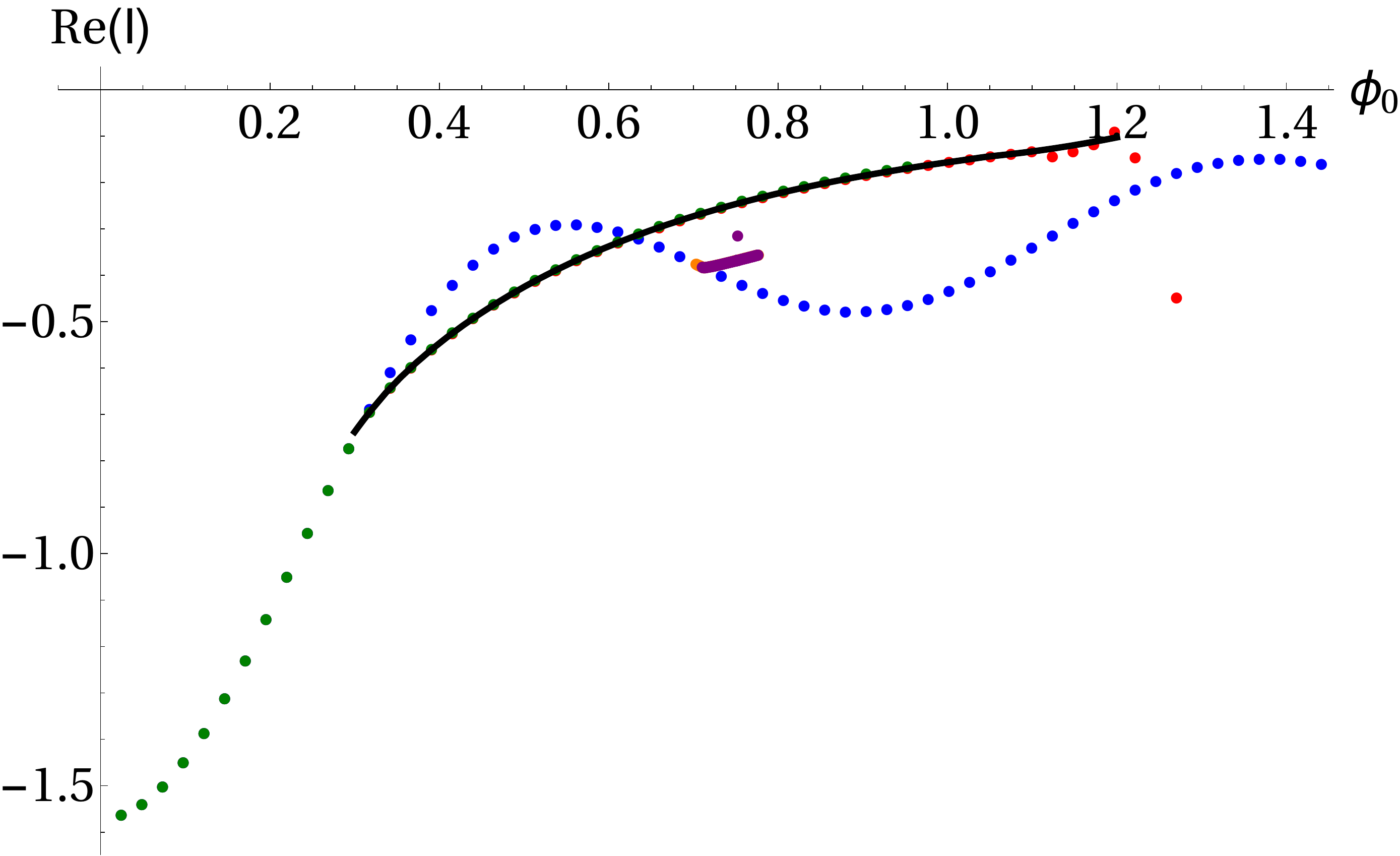}
\caption{$m=\sqrt{2}$} \label{fig:bifurcationActionmusqrt2}
\end{subfigure}
\begin{subfigure}[b]{0.32\textwidth}
\includegraphics[width=\textwidth]{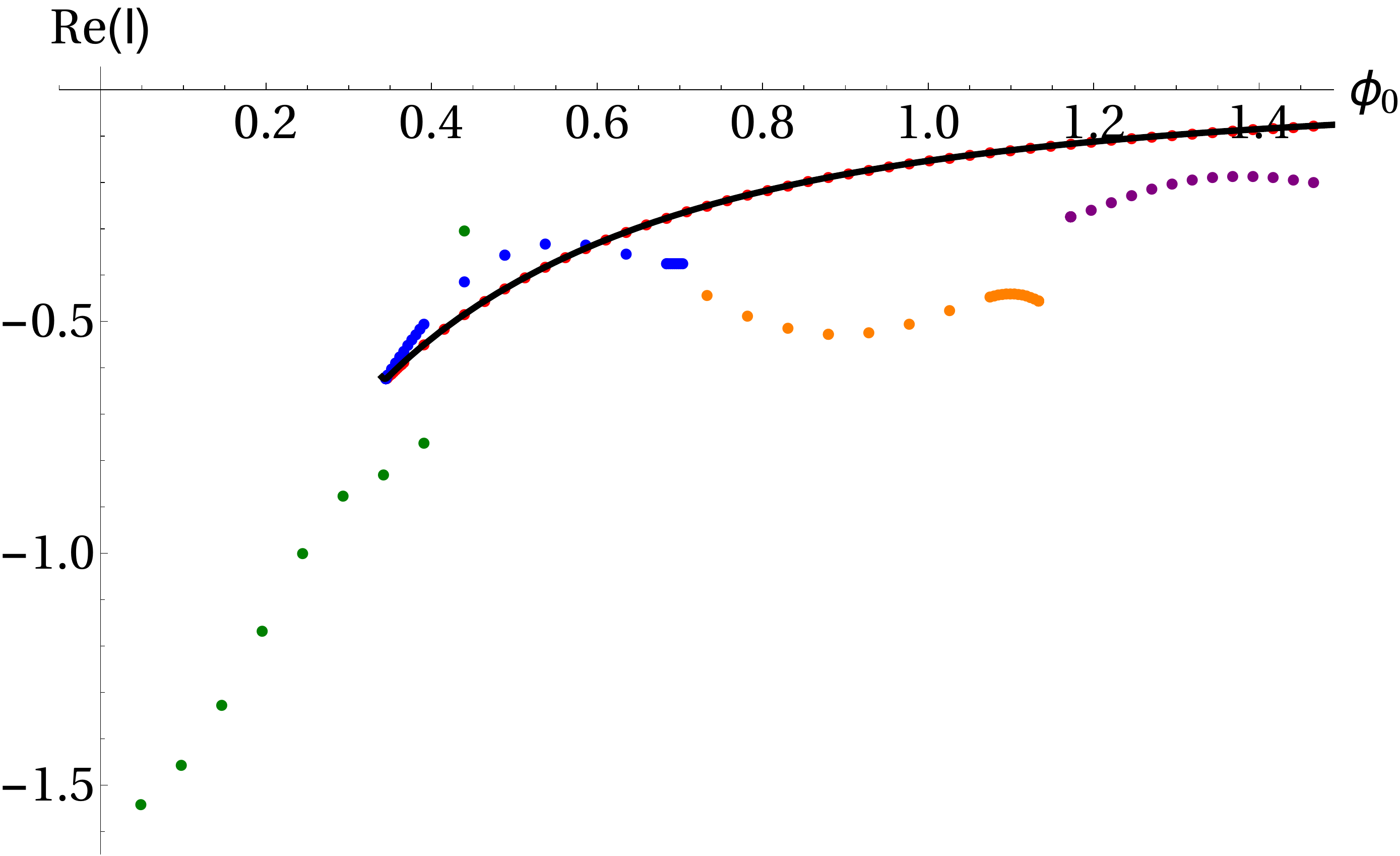}
\caption{$m=1.43$} \label{fig:bifurcationActionmu143}
\end{subfigure}
\caption{The real part of the action versus $\phi_0$ that specifies the relative probabilities of the solutions shown in Figure \ref{fig:bifurcation}. The colors coincide with the colors used there. For the case $m^2=2$ the red and the green branch coincide. The full black curve shows the solutions that obey both sets of classicality conditions.} \label{fig:bifurcationAction}
\end{figure}

The last figure of this section shows the relative probabilities of the different classical histories predicted by the NBWF, which are given by $\sim e^{-2\text{Re}(I)}$. In general the solution with the most negative real part of the on-shell action provides the dominant contribution to the wave function. In Figure \ref{fig:bifurcationAction} we show the real part of the action for each family of saddle points, whether classical or not. Notice that for $m^2=2$ the green and the red branch coincide perfectly as a consequence of the enhanced symmetry. Figure \ref{fig:bifurcationAction} shows that the action of the new solutions is of the same order of magnitude as the previously known solutions in red\footnote{The results shown in Figure \ref{fig:bifurcation} for the green branch of solutions with $m = 1.4$ are numerically unstable. We do not expect the apparent divergence to be physical, which should thus not be interpreted as a non-normalizable direction of the probability density.}. At first sight it seems non-trivial to decide which saddle point dominates for a given $\phi_0$.

However the new set of asymptotic classicality conditions selects a unique saddle point solution for each value of $\alpha$. Specifically Figure \ref{fig:classcondbeta} shows that the only classical saddle point solutions are on the red branch for sufficiently large $\phi_0$.
These solutions are denoted with a full black line in Figure \ref{fig:bifurcationAction}. Conditioning on asymptomatically classical behavior, therefore, restores the NBWF prediction of a one-parameter set of classical homogeneous isotropic universes with relative probabilities favoring a low amount of scalar field driven inflation.

\section{Conclusion}

We have derived a sufficient set of conditions on the Euclidean boundary theory in dS/CFT for it to predict classical, Lorentzian bulk evolution for large spatial volumes. The conditions amount to the requirement that the vevs corresponding to the external sources in the dual partition function must be approximately real. This in turn leads to a restriction on the values of the sources and on the path integral defining the partition functions if the dual theory is to be compatible with the asymptotic semiclassical structure implied by the WDW equation.

To derive the new set of classicality conditions, we first expressed the bulk wave function for large spatial volumes in terms of the sources of the dual partition function. This enabled us to put forward a sharper formulation of dS/CFT in which the wave function of the universe is directly related to the dual partition functions. 

The conditions under which the boundary theory predicts classical bulk evolution are stronger than the criteria usually employed in quantum cosmology. We illustrated this in a minisuperspace model comprising homogeneous isotropic histories in gravity coupled to a scalar field, where we identified several families of histories which are predicted to behave classically according to the old classicality conditions but do not obey the new conditions. Besides a number of exotic histories in which the scalar field is large, these also include histories which are relatively small perturbations of empty de Sitter in which the scalar field is small everywhere. This appears to be a generalization to light scalars of the prediction \cite{Hartle2008} in the Hartle-Hawking state for heavy scalars that empty de Sitter is an isolated point in the ensemble of asymptotically classical histories.

\section*{Acknowledgments}
We thank Dionysios Anninos, Adam Bzowski, Frederik Denef, James Hartle and Kristof Moors for useful conversations. YV would like to thank Columbia University for their hospitality during the last phases of this work. The work of TH and YV is supported in part by the Belgian National Science Foundation (FWO) grant G.001.12 Odysseus and by the European Research Council grant no. ERC-2013-CoG 616732 HoloQosmos. RM is supported by FWO-Flanders.


\appendix

\section{Canonical transformations of wave functions} \label{sec:QMtransfo}

The goal of this appendix is to explain how wave functions in terms of different variables are related to each other if these variables are connected by a canonical transformation.

In the most general case one would consider any even-dimensional phase space $\mathcal{M}$ with a symplectic two-form $J^{(2)}$. However, to simplify the discussion to its essence, we will only consider a two-dimensional phase space parametrized by $(x, p)$, with $J^{(2)} = \D p \wedge \D x$.

We will assume the dynamics is governed by a Hamiltonian $H(x,p)$ which has a discrete spectrum and a unique ground state with energy $E_0 = 0$. The ground state wave function can then be written as a path integral in Euclidean time
\begin{align}
\psi_0 (\fx) &= \lim_{\tau \to -\infty} \sum_{n \geq 0} e^{E_n \tau} \psi_n(\fx) \propto \lim_{\tau \to -\infty} \int_{x(\tau)=0}^{x(0)=\fx}{\cD x(\tau) \cD p(\tau)\ e^{i \int{p \D x} - \int{H(x,p) \D \tau}}}  \ .  \label{eq:groundState}
\end{align}
We have used the standard method of converting a transition amplitude into a path integral, familiar from calculations of transition amplitudes in real time $t = -i \tau$, i.e. by decomposing the unity operator: $1 = \int{\D x\ \ket{x} \bra{x}} = \int{\D p\ \ket{p} \bra{p}}$.

The same ground state can be described using any set of coordinates on phase space, $(X, P)$, which also form a Darboux pair $J^{(2)} = \D P \wedge \D X$. Such coordinates are related to $(x, p)$ by a canonical transformation, i.e. there exist functions 
\begin{align}
\xi: (X, P) &\mapsto \xi(X, P) = x   \ , &  \pi: (X, P) &\mapsto \pi(X, P) = p  \ , \label{eq:coordChange}
\end{align}
that map the $(X, P)$-coordinates of each point in phase space to its $(x, p)$-coordinates and that satisfy 
\begin{align}
\pi(X, P) \D \xi(X, P) &= P \D X + \D f(X, P)   \ . \label{eq:canonical}
\end{align}
Here we introduced $f$, the `generating function' of this canonical transformation. The goal of the remainder of this appendix is to show that the new wave function $\Psi_0(X)$, describing the same vacuum state, is related to the original wave function as
\begin{align}
\psi_0(\fx) &= \int{\D \fX\ e^{i \tilde{f}(\fX, \fx)} \Psi_0(\fX)}   \ ,    \label{eq:conclusion}
\end{align}
where $\tilde{f}(X, x)$ is defined as $f[X, \cP(X, x)]$, with $\cP$ solving the equation $\xi[X, \cP(X, x)] = x$.

To derive this relation carefully, we explicitly resolve the path integral measure as a dense set of ordinary integrals. 
In this discretized way, the path integral is
\begin{align}
\psi_0(\fx) &\sim \int{ \left( \prod_{n=0}^{N+1} \D x_n \right) \left( \prod_{n=0}^N \D p_n \right) \delta(\fx - x_{n+1}) e^{\sum_{n=0}^N \left[ i p_n (x_{n+1} - x_n) - H(x_i, p_i) \Delta \tau \right]} } \nonumber \\
&= \int{ \left( \prod_{n=0}^N \D x_n \D p_n \right) e^{i p_N (\fx - x_N) + i \sum_{n=0}^{N-1} i p_n (x_{n+1} - x_n) - \sum_{n=0}^N H(x_n, p_n) \Delta \tau } }  \ . \nonumber 
\end{align}
On each of these discretized slices, we can change the coordinates from ${x_n, p_n}$ to ${X_n, P_n}$ using \eqref{eq:coordChange}. Since the Jacobian of each of these transformations is 1, which is a general property of canonical transformations, we get
\begin{align}
\psi_0(\fx)  &\sim \int{ \left( \prod_{n=0}^N \D X_n \D P_n \right) e^{i \pi(X_N, P_N) [\fx - \xi(X_n, P_N)] - \sum_{n=0}^N H(X_n, P_n) \Delta \tau } } \nonumber \\
&\qquad\qquad\qquad\qquad\qquad \cdot e^{i \sum_{n=0}^{N-1} i \left[ P_n (X_{n+1} - X_n) + f(X_{n+1}, P_{n+1}) - f(X_n, P_n) \right]}   \label{eq:slowStart} \\
&= \int{ \left( \prod_{n=0}^N \D X_n \D P_n \right) e^{i \pi(X_N, P_N) [\fx - \xi(X_n, P_N)] + i f(X_N, P_N) + \sum_{n=0}^{N-1} \left[ i P_n (X_{n+1} - X_n) - H(X_n, P_n) \Delta \tau \right] } } \ . \nonumber
\end{align}
In these steps we have ignored terms proportional to $\Delta X_n \Delta \tau$ and $H(X_N, P_N) \Delta \tau$ since they will be irrelevant in the limit $\Delta \tau \to 0$.

From \eqref{eq:canonical} one can derive that for a canonical transformation,
\begin{align}
\frac{\partial f}{\partial P} &= \pi \frac{\partial \xi}{\partial P}.   \label{eq:canonicalIdentity}
\end{align}
If the functions $\xi$ and $\pi$ are linear in $P$, then $f$ will be at most quadratic. This implies that the $\D P_N$-integral in \eqref{eq:slowStart} is at most Gaussian, so we can solve it exactly by extremizing the exponent\footnote{If the transformation is not linear, the following step is still a good approximation whenever the method of steepest descent is accurate.}
\begin{align}
\frac{\partial \pi}{\partial P_N} [\fx - \xi(X_N, P_N)] - \pi \frac{\partial \xi}{\partial P_N} + \frac{\partial f}{\partial P_N} = 0   \ ,
\end{align}
where the last two terms cancel by virtue of \eqref{eq:canonicalIdentity}.

For a non-trivial canonical transformation $\partial \pi / \partial P \ne 0$, so we conclude that $P_N$ must solve the equation $\fx - \xi(X_N, P_N) = 0$, i.e. $P_N = \cP(X_N, \fx)$.

The above integral can thus be rewritten as
\begin{align}
\int{\D \fX\ e^{i f[\fX, \cP(\fX, \fx)]} \int{\left( \prod_{n=0}^N \D X_n \right) \left( \prod_{n=0}^{N-1} \D P_n \right) \delta(\fX - X_N) e^{\sum_{n=0}^{N-1} \left[ i P_n (X_{n+1} - X_n) - H(X_n, P_n) \Delta \tau \right] } } }  \ . \label{eq:finalStep}
\end{align}
The second integral is of the same form as the first line in \eqref{eq:slowStart}, i.e. (up to the limit $\tau \to -\infty$ and normalization) it will give the path integral $\Psi_0(\fX)$ upon taking the limit $\Delta \tau \to 0$. Thus, we have derived \eqref{eq:conclusion}.

\section{Symmetry enhancement and bifurcation} \label{subs:bifurcate}

Figure \ref{fig:bifurcation} resembles the classic picture of symmetry breaking of a pitchfork bifurcation, where the mass $m^2$ of the scalar field acts as the order parameter and where the symmetric solution is given by $m^2=2$. 
At this point of enhanced symmetry, the diagram seems to be invariant under the transformation
\begin{align}(x_{TP}, \theta) \to (\pi-x_{TP}, -\theta -\pi) \ . \label{eqn:symmtransfo} \end{align}
In this section we will show that this transformation indeed maps asymptotically real solutions to asymptotically real solutions for $m^2=2$.

Consider the scalar field initial value at the SP, $\phi=\phi_0 e^{i\theta}$ and apply the phase transformation of \eqref{eqn:symmtransfo}. This transforms $\phi$ to $-\phi^* = -\phi_0 e^{-i\theta}$. This suggests that the ``mirrored'' solutions are related to each other by complex conjugation of their initial data. Indeed, this also holds for the scale factor, since the complex conjugation of the no-boundary condition is just the no-boundary condition again.
One can see that complex conjugation maps solutions of the equations of motion to other solutions. Furthermore, there is no mixing in the equations between terms that are even and odd in $\phi$. This implies that mapping $\phi \to -\phi$ gives a new solution to the equations of motion\footnote{Indeed, the subtraction by $\pi$ in \ref{eqn:symmtransfo} is not essential. It serves merely to keep $\theta$ within the range $[0, \pi]$.}.

Now, we are interested in solutions that become real at late time. Hence we need to study the asymptotic solutions along the Lorentzian axis in the case $m^2=2$:
\begin{align}
\phi &= i \eta_0 e^{i(x_{TP}+i t)} \frac{\alpha}{\sqrt{2}\pi} \gamma^3 + \ldots   \ , &  a &= \frac{\gamma}{i \eta_0} e^{-i(x_{TP} +it)} + \ldots   \ .
\end{align}
Since we are calculating the wave function of a real de Sitter universe, $\phi$ and $a$ should asymptote to real values. This is possible if the phase of $i \eta_0 \frac{\alpha}{\sqrt{2}\pi} \gamma^3$ is $-ix_{TP}$ and that of $ \gamma / i \eta_0$ is $ix_{TP}$. By the $\theta$-transformation in \eqref{eqn:symmtransfo}, these phases will be mapped to $i (x_{TP} - \pi)$ and $-i x_{TP}-i \pi$, respectively. Thus we constructed a new solution of the equations of motion in the complex tau-plane,
\begin{align}
\phi &= -\lvert i \eta_0 \frac{\alpha}{\sqrt{2}\pi} \gamma^3 \rvert e^{i x_{TP}} e^{i \tau}	\ , &	a &= \lvert \frac{\gamma}{i \eta_0} \rvert e^{-i x_{TP}} e^{-i \tau}	\ .
\end{align}
Now one can see that there exists an asymptotic direction in which both of these become real: $\tau = \pi - x_{TP} + i t$, with $t \to \infty$.

Hence, for this value of the mass, we have obtained a set of new solution with asymptotically real values for the matter field and scale factor. When $m^2 \neq 2$, the symmetry of the solutions is broken because the phases of the exponential in the asymptotic fields change. This explains the pitchfork-like bifurcation in Figure \ref{fig:bifurcation}.

\bibliography{Literature}
\end{document}